\documentclass[a4paper,11pt]{article}
\pdfoutput=1

\usepackage{jheppub}
\usepackage{tikz}
\usepackage{verbatim}
\usetikzlibrary{shapes}
\usetikzlibrary{plotmarks}
\usepackage{tikz-3dplot}

\definecolor{light}{rgb}{0,1,1}

\title{Super Yang-Mills on Branched Covers and Weighted Projective Spaces}
\author[1]{Roman Mauch}
\author[2]{and Lorenzo Ruggeri}
\affiliation[1]{Department of Physics and Astronomy, Uppsala Universitet, 752 37 Uppsala, Sweden}
\affiliation[2]{Yau Mathematical Sciences Center, Tsinghua University, Beijing, 100084, China}
\emailAdd{roman.mauch@physics.uu.se}
\emailAdd{ruggeri@mail.tsinghua.edu.cn}
\preprint{UUITP-10/24}

\abstract{
In this work we conjecture the Coulomb branch partition function, including flux and instanton contributions, for the $\mathcal{N}=2$ vector multiplet on weighted projective space $\mathbb{CP}^2_{\boldsymbol{N}}$ for equivariant Donaldson-Witten and ``Pestun-like'' theories. We claim that this partition function agrees with the one obtained from dimensional reduction of the 5d $\mathcal{N}=1$ vector multiplet on a certain branched cover of $S^5$. More precisely, the branch locus and indices have to be such that they match the singular locus and deficit angles in $\mathbb{CP}^2_{\boldsymbol{N}}$. Our conjecture is substantiated by checking that partition functions on spindles are similarly obtained from dimensional reduction of the 3d $\mathcal{N}=2$ vector multiplet on branched covers of $S^3$.
This work paves the way for obtaining partition functions on more generic symplectic toric orbifolds.
}

\begin{document}

\maketitle
\flushbottom

\section{Introduction}

    Since the early days of supersymmetric localisation \cite{Pestun:2007rz,Pestun:2016zxk} the study of observables for supersymmetric quantum field theories (SQFTs) on spaces with singularities has attracted much attention. Most notably, the supersymmetric Renyi entropy for $\mathcal{N}=2$ superconformal field theories (SCFTs) has been computed from the partition function of the SCFT on certain branched covers of $S^3$ \cite{Nishioka:2013haa}. This result was later generalised to branched covers of spheres $S^d$, for $d=2,4,5$ \cite{Crossley:2014oea,Mori:2015bro,Alday:2014fsa,Hama:2014iea}. While these spaces have conical surplus angles at the branch locus, recently there has been considerable interest in computing observables on spaces with orbifold singularities (i.e. conical deficit angles) instead. These theories, which are interesting in their own right, offer the possibility to holographically reproduce accelerating black hole solutions in supergravity with orbifold singularities in their near horizon geometry \cite{Ferrero:2020laf,Ferrero:2020twa,Ferrero:2021etw}. Recently, the authors of \cite{Inglese:2023wky,Inglese:2023tyc} computed the partition function on $\mathbb{CP}^1_{\boldsymbol{N}}\times S^1$ which was then used in \cite{Colombo:2024mts} to reproduce the entropy function of supersymmetric accelerating black holes in AdS$_4$. The computation was performed for two different theories denoted twist and anti-twist, which reduce on $S^2\times S^1$ to, respectively, the topologically twisted index \cite{Benini:2015noa} and the superconformal index \cite{Kim:2009wb,Imamura:2011su,Kapustin:2011jm}. Their result was obtained employing an orbifold index theorem \cite{10.1215/S0012-7094-96-08226-5}. 
   
    In this work we draw a connection between the aforementioned results on branched covers of $S^3$ and the spindle $\mathbb{CP}^1_{\boldsymbol{N}}$, which we subsequently extend to branched covers of $S^5$ and weighted projective space $\mathbb{CP}^2_{\boldsymbol{N}}$. Our starting point are existing results for the perturbative partition function of the $\mathcal{N}=2$ vector multiplet on a branched three-sphere $S^3_{\boldsymbol{\alpha}}$ \cite{Klebanov:2011uf,Nishioka:2013haa} and $\mathcal{N}=1$ vector multiplet on a branched five-sphere $S^5_{\boldsymbol{\alpha}}$ \cite{Hama:2014iea,Alday:2014fsa}\footnote{In this work we consider generic integer values for $\boldsymbol{\alpha}=(\alpha_1,\dots,\alpha_r)$, $r=2,3$ (cf. Renyi entropy).}. Both theories are in an $\Omega$-background, i.e. the supercharge squares (among other symmetries) to translations along a Killing vector field. The branched spheres admit locally free $S^1$-actions and our aim is to compute the partition function on the quotient by such actions (which we denote by $B^1_{\boldsymbol{\alpha}}$ and $B^2_{\boldsymbol{\alpha}}$ in the following). The method we use for the circle reduction is the one of \cite{Lundin:2021zeb,Lundin:2023tzw}: first, we consider quotients by a finite subgroup, $S^{2r-1}_{\boldsymbol{\alpha}}/\mathbb{Z}_h$ (for $r=2,3$). This introduces extra topological sectors in the partition function corresponding to non-trivial flat connections in the localisation locus. The circle reduction is then obtained in the limit $h\to\infty$, in which these topological sectors label flux through the two-cycle on the base $B^{r-1}_{\boldsymbol{\alpha}}$. We consider two different $S^1$-quotients with respect to a fixed choice of Killing vector: a Hopf-like quotient, leading to an equivariant Donaldson-Witten theory on the base and an anti-Hopf-like quotient, leading to a ``Pestun-like'' theory. As the main result, we compute the one-loop determinant and classical part of the full partition function on $B^{r-1}_{\boldsymbol{\alpha}}$ in \eqref{eq.1loop.spindle.top}-\eqref{eq.S2cl}, resp. \eqref{eq.CP2alphatop}-\eqref{eq.S4cl}.

    We observe that for $r=2$ our result for the one-loop determinant in each flux sector agrees, for both topological and exotic theories, with, respectively, the twist and anti-twist theories \cite{Inglese:2023wky} on the spindle upon the identification of weights $N_1=\alpha_1$, $N_2=\alpha_2$. In 4d, our result for the topologically twisted theory matches its counterpart on $\mathbb{CP}^2_{\boldsymbol{N}}$ \cite{Martelli:2023oqk} upon suitable identification of weights\footnote{A similar relation has been observed for 3d and 4d SCFTs on orbifolds and branched covers of spheres with a single non-trivial branch index in \cite{Nishioka:2016guu}.}. Guided by these observations, we make the \autoref{conjecture} that also the one-loop determinant for the exotically twisted $\mathcal{N}=2$ vector multiplet on $\mathbb{CP}^2_{\boldsymbol{N}}$ is simply given by our result on $B^2_{\boldsymbol{\alpha}}$ for each flux sector. Moreover, we believe that this matching for spaces with singular and branch locus works more generally. It would therefore be interesting to compute the one-loop determinant also for the $S^1$-reduction from branched covers of other toric Sasakian 5-manifolds. We expect the results to match with one-loop determinants of the corresponding symplectic toric 4-orbifolds \cite{Martelli:2023oqk,Colombo:2023fhu}.

    Finally, capitalising on equivariance of our setup under a $T^2$-action, for the theory on $B^2_{\boldsymbol{\alpha}}$ we factorise the one-loop determinant in each flux sector into individual contributions from the torus fixed points. In each factor, the Coulomb branch parameter is shifted in a certain way by the flux through the two-cycle. The instanton part of the partition function is commonly obtained as a product of Nekrasov partition functions over the fixed points \cite{Nekrasov:2003vi,Bershtein:2015xfa,Festuccia:2018rew,Festuccia:2019akm,Lundin:2023tzw}, where in each factor the Coulomb branch parameter is shifted in precisely the way obtained from factorisation of the one-loop determinant. Provided the equivalence of one-loop determinants on $B^2_{\boldsymbol{\alpha}}$ and $\mathbb{CP}^2_{\boldsymbol{N}}$ holds in each flux sector, their shifts also agree. With the Nekrasov partition function on a neighbourhood $\mathbb{C}^2/\mathbb{Z}_{N_i}$ of the $i$th fixed point being known \cite{Fucito:2004ry,Bonelli:2011jx,Bonelli:2011kv,Bonelli:2012ny,Bruzzo:2013nba,Bruzzo:2013daa}, we conclude our work by providing the (integrand of the) full partition function \eqref{eq--7.full}, including classical, one-loop and instanton part at all flux sectors for the $\mathcal{N}=2$ vector multiplet on weighted projective space $\mathbb{CP}^2_{\boldsymbol{N}}$ both, for topological and exotic theories.

    The outline is as follows. In \autoref{sec--2} we introduce the geometric setup of branched spheres $S^{2r-1}_{\boldsymbol{\alpha}}$ and discuss the localisation computation of the vector multiplet on a branched three- and five-sphere.
    \hyperref[sec--3]{Section 3} describes the dimensional reduction of the theory along two different locally free $S^1$-directions which yield the one-loop determinant on the respective quotient $S^{2r-1}_{\boldsymbol{\alpha}}/S^1$ for a fixed flux sector. For $r=2,3$ we compute these explicitly in \autoref{sec--4} and \ref{sec--5}. Using equivariance of the setup, we factorise the one-loop determinants into contributions from each torus fixed point in \autoref{sec--6}. Based on our observation that the determinants on $S^{3}_{\boldsymbol{\alpha}}/S^1$ match the ones on weighted projective space and this also holds for the topological twist on $S^{5}_{\boldsymbol{\alpha}}/S^1$, in \autoref{sec--7} we conjecture and give some arguments in favour of such a match also for the exotic theory. Finally, in \autoref{app--geometry} we provide more details on the geometric setup and in \autoref{app--A} we compute the small radius limit of the spindle index.

\section{Branched Spheres}\label{sec--2}

    In this section we introduce our geometric setup and field theoretic starting point, the $\mathcal{N}=2$ (resp. $\mathcal{N}=1$) vector multiplet partition function on branched spheres in three (resp. five) dimensions. These are existing results from \cite{Klebanov:2011uf,Nishioka:2013haa} (resp. \cite{Hama:2014iea,Alday:2014fsa}) to which we refer the reader for more detail\footnote{In 3d the partition function was obtained by first computing it on a background where the singularities are resolved (which is a $Q$-exact deformation of the theory on $S^3_{\boldsymbol{\alpha}}$) and then taking the singular limit. The result agrees with the one on a squashed sphere with squashings reciprocal to the $\{\alpha_i\}$. A similar relation has been found in 5d for $\boldsymbol{\alpha}=(\alpha_1,1,1)$ (which is easily extended to $\alpha_2,\alpha_3\neq1$) and also for a squashed $S^5_{\boldsymbol{\alpha}}$.}.

    Throughout this article, we always consider cohomologically twisted theories with simply-connected gauge group $G$ (see \cite{Kallen:2011ny,Kallen:2012cs} for the definition of cohomological fields and the corresponding cohomological complex in three and five dimensions). For both, the (squashed or unsquashed) spheres $S^3,S^5$ and (the resolution of) their branched cover, Killing spinors solving the corresponding rigid supergravity background are known \cite{Nishioka:2013haa,Alday:2014fsa,Hama:2014iea} and hence, one can switch between the physical and twisted theories bijectively.

    \subsection{Geometric Setup}
    
        The branched spheres in consideration, which we always denote by $S^{2r-1}_{\boldsymbol{\alpha}}$ in the following (for $r=2,3$), are obtained from the ordinary ones simply by extending the periodicity of the azimuthal angles of $S^{2r-1}$ to an integer multiple of $2\pi$. If we denote said angles by $\theta_i$, $i=1,\dots,r$ with $\theta_i\sim\theta_i+2\pi$ then we extend these to new angles $\tilde{\theta}_i$ with $\tilde{\theta}_i\sim\tilde{\theta}_i+2\pi\alpha_i$, where $\alpha_i\in\mathbb{Z}$. The integers $\{\alpha_i\}$ are called the \textit{branch indices} and in the notation $S^{2r-1}_{\boldsymbol{\alpha}}$ are collected in the subscript $\boldsymbol{\alpha}=(\alpha_1,\dots,\alpha_r)$. The branched sphere is an $\alpha_1\dots\alpha_r$-fold branched cover of the ordinary sphere (see \autoref{app--geometry} for more detail on the branching structure). More generally, we obtain a squashed branched sphere by extending the periodicities as above, but for a squashed $S^{2r-1}$ instead.
        
        \paragraph{Three Dimensions.} The metric on $S^3_{\boldsymbol{\alpha}}$ can be obtained from the round one and in terms of the canonical angles reads
        \begin{align}
            \dd s^2=\dd\phi^2+\alpha_1^2\sin^2\phi\dd\theta_1^2+\alpha_2^2\cos^2\phi\dd\theta_2^2,
        \end{align}
        with $\phi\in[0,\pi/2]$. It was shown in \cite{Nishioka:2013haa} that this space has conical singularities in the curvature, supported on the Hopf-link comprised of the two $S^1$-subspaces $\phi=0,\pi/2$, with surplus angles $\alpha_1,\alpha_2$. The two locally free $S^1$-actions we are interested in (which are Killing) are given by
        \begin{equation}\label{eq--2.free.vector.3d}
            \X=\partial_{\tilde{\theta}_1}\pm\partial_{\tilde{\theta}_2}=\frac{1}{\alpha_1}\partial_{\theta_1}\pm\frac{1}{\alpha_2}\partial_{\theta_2}.
        \end{equation}
        $S^3_{\boldsymbol{\alpha}}$ can be viewed as a ``branched'' (anti-)Hopf fibration
        \begin{equation}
            \begin{tikzcd}[column sep=20pt]
                S^1\ar[r,hook] & S^3_{\boldsymbol{\alpha}}\ar[r] & B^1_{\boldsymbol{\alpha}}
            \end{tikzcd}
        \end{equation}
        along this action, whose base $B^1_{\boldsymbol{\alpha}}=S^3_{\boldsymbol{\alpha}}/S^1$ is, in fact, simply the two-sphere\footnote{Despite this fact, we stick with the notation $B^1_{\boldsymbol{\alpha}}$ for the base space, signifying that the resulting partition function on $B^1_{\boldsymbol{\alpha}}$ is \textit{not} the one on $S^2$.}. We can also reach $B^1_{\boldsymbol{\alpha}}$ by first considering finite quotients\footnote{\label{fn--gcd}For now, we take $\gcd(h,\alpha_i)=1$, $h>\alpha_i$ for all $i=1,\dots,r$. This can still be satisfied in the limit where $h$ becomes large by virtue of Euclid's theorem. See \autoref{app--geometry} for a discussion of other choices.} $S^3_{\boldsymbol{\alpha}}/\mathbb{Z}_h$ along \eqref{eq--2.free.vector.3d}, giving branched lens spaces which we denote $L^3_{\boldsymbol{\alpha}}(h,\pm1)$. Note that the finite quotient introduces holonomies classified by $\mathbb{Z}_h$. We arrive at the base space $B^1_{\boldsymbol{\alpha}}$ by taking the limit $h\to\infty$ (see \cite{Lundin:2023tzw}).

        \paragraph{Five Dimensions.} The metric on $S^5_{\boldsymbol{\alpha}}$ can be obtained from the round one and in terms of the canonical angles reads
        \begin{equation}\label{eq--2.metric.5d}
            \dd s^2=\dd\varphi^2+\sin^2\!\varphi\,\dd\phi+\alpha_3^2\cos^2\!\varphi\dd\theta_3^2+\alpha_2^2\sin^2\!\varphi\sin^2\!\phi\,\dd\theta_2^2+\alpha_1^2\sin^2\!\varphi\cos^2\!\phi\,\dd\theta_1^2,
        \end{equation}
        with $\varphi,\phi\in[0,\pi/2]$. Similar to the three-dimensional case, the curvature can be shown to have conical singularities on the three (branched) $S^3$-subspaces $\varphi=\pi/2$ (with surplus $2\pi\alpha_3$), $\phi=0$ (with surplus $2\pi\alpha_2$) and $\phi=\pi/2$ (with surplus $2\pi\alpha_1$). The two locally free $S^1$-actions (whose generating directions are Killing) of interest are:
        \begin{equation}\label{eq--2.free.vector.5d}
            \X=\frac{1}{\alpha_1}\partial_{\theta_1}+\frac{1}{\alpha_2}\partial_{\theta_2}\pm\frac{1}{\alpha_3}\partial_{\theta_3}.
        \end{equation}
        Similar to the three-dimensional case, we want to take the $S^1$-quotient by this action, $B^2_{\boldsymbol{\alpha}}=S^5_{\boldsymbol{\alpha}}/S^1$ which is, in fact, simply $\mathbb{CP}^2$. 
        Again, as an intermediate step we can take finite quotients\footnoteref{fn--gcd} $S^5_{\boldsymbol{\alpha}}/\mathbb{Z}_h$ giving (generalised) branched lens spaces $L^5_{\boldsymbol{\alpha}}(h,\pm1)$ with extra topological sectors classified by $\mathbb{Z}_h$.

    \subsection{Localisation}\label{subsec--localisation}


        The theories we want to place on $S^3_{\boldsymbol{\alpha}}$ (resp. $S^5_{\boldsymbol{\alpha}}$) are those of the $\mathcal{N}=2$ (resp. $\mathcal{N}=1$) vector multiplet. We again refer to \cite{Klebanov:2011uf,Nishioka:2013haa} (resp. \cite{Hama:2014iea,Alday:2014fsa}) for supersymmetry backgrounds and actions.

        For localisation, note that away from the branch locus, locally the branched sphere is diffeomorphic to the ordinary sphere and hence the computation is locally the same. At the branch locus, however, $S^{2r-1}_{\boldsymbol{\alpha}}$ is singular. This has the following consequences: (i) the background R-symmetry connection $A_R$ is flat except at the branch locus where it has a singular field strength compensating for the curvature singularity and (ii) boundary conditions\footnote{The term ``boundary conditions'' here stems from the fact that there exists a conformal mapping from the branched $(2r-1)$-dimensional sphere with $\boldsymbol{\alpha}=(\alpha,1,\dots,1)$ to $\mathbb{H}^{r-1}\times S^1$ where the branch locus is mapped to the circle at infinity, i.e. is an actual boundary. One could, for instance, consider dynamical gauge fields that asymptotically have non-trivial holonomy around the $S^1$ \cite{Hosseini:2019and}. However, here we restrict to trivial holonomies for the dynamical fields.} have to be imposed for the fields of the theory along the branch locus. 

        Computing the one-loop partition function from localisation essentially reduces to counting (normalisable) holomorphic functions weighted by their $T^r$-action (see e.g. \cite{Qiu:2016dyj}). On $S^{2r-1}=\{(z_1,\dots,z_r)\in\mathbb{C}^r|\,|z_1|^2+\dots+|z_r|^2=1\}$ a basis for such maps is given by polynomials in $z_1,\dots,z_r$. When we move to the branched cover by extending the periodicities as discussed in the previous subsection, such a basis is instead given by $z_1^{1/\alpha_1},\dots,z_r^{1/\alpha_r}$ (where $\arg z_i^{1/\alpha_i}=\tilde{\theta}_i/\alpha_i$) and the boundary conditions on the fields are that the holomorphic functions be \textit{smooth in $\{z_i^{1/\alpha_i}\}$} rather than $\{z_i\}$. Consequently, when expanding in Fourier modes $\exp(\ii n_1\theta_1+\dots+\ii n_r\theta_r)$, the charge under the locally free $S^1$-action \eqref{eq--2.free.vector.3d}, resp. \eqref{eq--2.free.vector.5d} is $\frac{n_1}{\alpha_1}+\dots+\frac{n_{r-1}}{\alpha_{r-1}}\pm\frac{n_r}{\alpha_r}$ with $n_1,\dots,n_r\in\mathbb{Z}$. 
        
        The locally free $S^1$-action can be parametrised by an angle $\xi$ with $\xi\sim\xi+2\pi\lcm\boldsymbol{\alpha}$. Hence, \textit{locally}, smooth functions on the branched cover can be expanded in terms of modes $\exp(\ii t\xi/\lcm\boldsymbol{\alpha})$ along the $S^1$ and the charge $t\in\mathbb{Z}$ is related to the ones corresponding to the canonical angles by
        \begin{equation}\label{eq--2.tlcm}
            \frac{t}{\lcm\boldsymbol{\alpha}}=\frac{n_1}{\alpha_1}+\dots+\frac{n_{r-1}}{\alpha_{r-1}}\pm\frac{n_r}{\alpha_r}.
        \end{equation} 
        This relation will be applied when reducing along the locally free $S^1$ later on.

        \paragraph{Perturbative Partition Function.}
            Our observations above suggest that the result of the one-loop computation on a squashed $S^{2r-1}_{\boldsymbol{\alpha}}$ with squashings $\{\omega_i\}_{i=1,\dots,r}$ is simply that on $S^{2r-1}$ \cite{Hama:2011ea,Imamura:2011wg,Kallen:2012va,Imamura:2012efi,Lockhart:2012vp} for a background with Killing vector $\boldsymbol{\R}=(\frac{\omega_1}{\alpha_1},\dots,\frac{\omega_r}{\alpha_r})$ and yields the multiple sine function
            \begin{equation}\label{eq.multiple.sine}
                S_r(\ii \alpha(a)|\boldsymbol{\R})=\prod_{\boldsymbol{n}\in\mathbb{Z}^r_{\geq 0}}(\boldsymbol{n}\cdot\boldsymbol{\R}+ \ii\alpha(a) )\prod_{\boldsymbol{n}\in\mathbb{Z}^r_{>0}}(\boldsymbol{n}\cdot\boldsymbol{\R}- \ii\alpha(a) )^{(-1)^{r-1}}
            \end{equation}
            for a fixed root $\alpha$ of the gauge group $G$ and Coulomb branch parameter $a$ in the Cartan subalgebra of $G$. Note that for non-trivial squashing the supercharge used for localisation does not square to rotations along the locally free direction $\X$ (cf. footnote \ref{fn--Q.square}). The perturbative partition function is obtained as
            \begin{equation}\label{eq--2.pert.partition}
                \mathcal{Z}^\text{pert}_{S^{2r-1}_{\boldsymbol{\alpha}}}=\int_{\mathfrak{h}}\dd a~\e{-S^\text{cl}_{2r-1}}\det{}^{\prime}_{\text{adj}}~S_r(\ii\alpha(a)|\boldsymbol{\R}),
            \end{equation}
            where $\mathfrak{h}\subset\mathfrak{g}$ denotes the Cartan subalgebra and the determinant is over the roots of $G$, where the prime signifies that possible zero-mode contributions have already been canceled by the Vandermonde determinant (which arises when restricting integration from $\mathfrak{g}$ to $\mathfrak{h}$). The classical action $S^\text{cl}_{2r-1}$ for $r=2,3$ is given by
            \begin{align}\label{eq.3d.classical.Salpha}
                S^\text{cl}_3=-\frac{\ii\pi k}{\R_1\R_2}\tr a^2,\qquad\qquad
                S^\text{cl}_5=\frac{(2\pi)^3}{g^2_\text{YM}\R_1\R_2\R_3}\tr a^2
            \end{align}
            with Chern-Simons level $k$, Yang-Mills coupling $g^2_\text{YM}$ and the weights $\R_i=\frac{\omega_i}{\alpha_i}$. We comment on instanton contributions to the full partition function in \autoref{sec--7}. 
            
            Finally, we impose that the squashing leave the inner product $\X\cdot\R$ invariant, i.e. 
            \begin{equation}\label{eq.squashingcontraint}
                \begin{aligned}
                    \text{top:}\quad&\omega_1+\dots+\omega_r=r,\\
                    \text{ex:}\quad&\omega_1+\dots+\omega_{r-1}-\omega_r=r-2,
                \end{aligned}
            \end{equation}
            for the two choices of $\X$ corresponding to a Hopf (resp. anti-Hopf) fibration. We will refer to the respective $(2r-2)$-dimensional theory on $S^{2r-1}_{\boldsymbol{\alpha}}/S^1$ as the topological (resp. exotic) theory\footnote{We adopt this terminology from \cite{Festuccia:2019akm}. See also the comment at the end of \autoref{subsec--reduction}.}, hence the labels \textit{top} and \textit{ex} in \eqref{eq.squashingcontraint}.

\section{Partition Function on \texorpdfstring{$S^{2r-1}_{\boldsymbol{\alpha}}/S^1$}{S²ʳ⁻¹ₐ/S¹}}\label{sec--3}
    
    Having the result for the perturbative partition function on $S^{2r-1}_{\boldsymbol{\alpha}}$ at hand, in this section we will perform the dimensional reduction along the two\footnote{The two $S^1$-directions differ by the relative direction with respect to the choice of Killing vector $\R$. However, in this section we discuss the general procedure of the reduction and will mostly not distinguish between the two choices of locally free $S^1$ explicitly (cf. end of \autoref{subsec--reduction}).} locally free $S^1$-directions identified in \eqref{eq--2.free.vector.3d}, resp. \eqref{eq--2.free.vector.5d}. In particular, this will produce the one-loop contribution of the vector multiplet on the quotient space $B^{r-1}_{\boldsymbol{\alpha}}$ in all flux sectors (note $H^2(B^{r-1}_{\boldsymbol{\alpha}})\simeq\mathbb{Z}$).

    The quotient procedure follows the method in \cite{Lundin:2023tzw}, to which we refer for a detailed explanation. We first consider the theory on finite quotients $S^{2r-1}_{\boldsymbol{\alpha}}/\mathbb{Z}_h$, introduced in the previous section. This requires to extend the localisation locus to include flat connections with non-trivial holonomy around the locally free $S^1$. This, in turn, gives a one-loop contribution around such flat connections to the partition function on generalised lens spaces $L^{2r-1}_{\boldsymbol{\alpha}}(h,\pm1)$. In the limit $h\to\infty$, we reach $B^{r-1}_{\boldsymbol{\alpha}}$ and the contributions from non-trivial flat connections turn into those with non-trivial flux. Note that, unlike standard Kaluza-Klein reduction, in our case, none of the higher KK modes are integrated out, but rather distribute themselves into different topological sectors on the base.

    \subsection{Finite Quotients}

        In order to understand how the partition function \eqref{eq--2.pert.partition} has to be modified for a finite quotient of the underlying space, it suffices to analyse how the fields contributing to the one-loop determinant change. As was mentioned in \autoref{subsec--localisation}, these are precisely the holomorphic functions on the space. 
        
        Holomorphic functions on $S^{2r-1}_{\boldsymbol{\alpha}}$ are in one-to-one correspondence with points $\boldsymbol{n}$ in the integral cone $\mathbb{Z}^r_{\ge0}$ labelling their $T^r$-charge. We can decompose the cone into slices corresponding to functions with fixed charge $t$ under $\X$, using the relation \eqref{eq--2.tlcm}:
        \begin{equation}\label{eq.sliceBt}
            \mathcal{B}_{t,\boldsymbol{\alpha}}=\left\{(n_1,\dots,n_{r-1})\in\mathbb{Z}^{r-1}_{\ge0}\Big|\exists\, n_r\in\mathbb{Z}_{\ge0}:\tfrac{t}{\lcm\boldsymbol{\alpha}}=\tfrac{n_1}{\alpha_1}+\dots+\tfrac{n_{r-1}}{\alpha_{r-1}}\pm\tfrac{n_r}{\alpha_r}\right\}.
        \end{equation}
        Note that in the exotic theory $t\in\mathbb{Z}$ and the slices can be non-compact while in the topological one $t\in\mathbb{Z}_{\ge0}$ and the slices are always compact. Now we can rewrite the product over the cone $\mathbb{Z}^r_{\ge0}$ in the multiple sine function as
        \begin{equation}\label{eq.tperturbative.Salpha}
            S_r(\ii\alpha(a)|\boldsymbol{\R})=\prod_t\prod_{\boldsymbol{n}\in\mathcal{B}_{t,\boldsymbol{\alpha}}}\left(\boldsymbol{n}\cdot\tilde{\boldsymbol{\R}}\pm\frac{\omega_r t}{\lcm\boldsymbol{\alpha}}+\ii\alpha(a)\right)\prod_{\boldsymbol{n}\in\mathring{\mathcal{B}}_{t,\boldsymbol{\alpha}}}\left(\boldsymbol{n}\cdot\tilde{\boldsymbol{\R}}\pm\frac{\omega_r t}{\lcm\boldsymbol{\alpha}}-\ii\alpha(a)\right)^{(-1)^{r-1}},
        \end{equation}
        where $\mathring{\mathcal{B}}_{t,\boldsymbol{\alpha}}$ denotes the interior of the slice \eqref{eq.sliceBt} and $\tilde{\boldsymbol{\R}}=(\frac{\epsilon_1}{\alpha_1},\dots,\frac{\epsilon_{r-1}}{\alpha_{r-1}})$ the weights of the residual $T^{r-1}$-action on the slice, with
        \begin{equation}\begin{split}\label{eq.epsilon.parameters}
                \text{top:}&\quad\epsilon_i=\omega_i-\omega_r,\\
                \text{ex:}&\quad\epsilon_i=\omega_i+\omega_r,
        \end{split}\end{equation}
        for $i=1,\dots,r-1$. Note that in the unsquashed limit, $\omega_i\to1$ for all $i=1,\dots,r$, the equivariance parameters $\{\epsilon_i\}$ vanish\footnote{\label{fn--Q.square}This is expected since in the unsquashed limit the supercharge in the $(2r-1)$-dimensional theory squares precisely to rotations along the locally free $S^1$-direction, $Q^2=\mathcal{L}_\X$. Accordingly, once we take the quotient the supercharge squares to zero and equivariance is lost.} in the topological case, while they stay finite for the exotic case. Finally, we can write $\omega_r=\omega_r(\epsilon_1,\dots,\epsilon_{r-1})$ by virtue of \eqref{eq.squashingcontraint}.
        We point out that \eqref{eq.tperturbative.Salpha} for topological and exotic case are simply two equivalent rewritings of \eqref{eq.multiple.sine}.

        \paragraph{Taking the Quotient.}
            Naively, only holomorphic functions of charge $t=0\mmod h$ along $\X$ descend to the quotient. However, note that $L^{2r-1}_{\boldsymbol{\alpha}}(h,\pm1)$ has extra topological sectors labelled by $\mathbb{Z}_h$ in the fundamental group. These are in bijection with flat connections of line bundles over $L^{2r-1}_{\boldsymbol{\alpha}}(h,\pm1)$ (up to gauge transformations). By standard arguments, on the localisation locus the gauge group is broken to its maximal torus $U(1)^{\rk G}$ \cite{Blau:1994rk} and flat connections in the locus are characterised (up to gauge transformations) by representations $\Hom(\mathbb{Z}_h,U(1)^{\rk G})$ (up to conjugation). These can be labelled by certain elements $\mathfrak{m}$ in the Cartan subalgebra of $G$ (e.g., for $G=U(N)$ we have $\mathfrak{m}=\diag(m_1,\dots,m_N)\in\mathbb{Z}_h^{N\times N}$, $m_i\le m_{i+1}$).
            Hence, flat line bundles become part of the locus and apart from functions with $t=0\mmod h$ (which descend to \textit{functions} on $L^{2r-1}_{\boldsymbol{\alpha}}(h,\pm1)$), for fixed $\mathfrak{m}$ we should also allow for ones of charge
            \begin{equation}\label{eq.projection}
                t=\alpha(\mathfrak{m})\mmod h
            \end{equation}
            which descend to \textit{sections} of the corresponding flat line bundle on $L^{2r-1}_{\boldsymbol{\alpha}}(h,\pm1)$. From these observations we conclude that the finite quotient along $\X$ can be implemented at the level of the one-loop determinant simply by restricting the product in \eqref{eq.tperturbative.Salpha} to $t$ such that \eqref{eq.projection} is satisfied. The partition function then becomes
            \begin{equation}\label{eq.fullL3}
                \mathcal{Z}_{L^{2r-1}_{\boldsymbol{\alpha}}(h,\pm1)}=\sum_{\mathfrak{m}}\int_{\mathfrak{h}}\dd a~\e{-S^\text{cl}_{2r-1}/h}\det{}^\prime_\text{adj}~S_r(\ii\alpha(a)|\boldsymbol{\R})|_{t=\alpha(\mathfrak{m})\mmod h}
            \end{equation} 
            with $S^\text{cl}_5$ as in \eqref{eq.3d.classical.Salpha} but 
            \begin{equation}\label{eq.classicalL3}
                S^\text{cl}_3=-\frac{\ii\pi k}{\R_1\R_2}\tr(a^2\pm\mathfrak{m}^2)
            \end{equation}
            due to the Chern-Simons term \cite{Guadagnini:2017lcz}. Note that now the topological and exotic choice describe different theories for $h>1$.

    \subsection{Reduction to the Base}\label{subsec--reduction}
            
        We arrive at the quotient space $B^{r-1}_{\boldsymbol{\alpha}}=S^{2r-1}_{\boldsymbol{\alpha}}/S^1$ by taking the limit $h\to\infty$ where the $S^1$-fibre shrinks to a point (a formal treatment of this limit for five-manifolds can be found in \cite{Lundin:2023tzw}, appendix A). For $r=2$, the theory for the $\mathcal{N}=2$ vector multiplet on $S^3_{\boldsymbol{\alpha}}$ descends to the one for an $\mathcal{N}=(2,2)$ vector multiplet on $B^1_{\boldsymbol{\alpha}}$. Similarly, for $r=3$, the $\mathcal{N}=1$ vector multiplet on $S^5_{\boldsymbol{\alpha}}$ descends to an $\mathcal{N}=2$ vector multiplet on $B^2_{\boldsymbol{\alpha}}$. Depending on whether we reduce along the topological or exotic direction, we obtain two differently twisted theories on $B^{r-1}_{\boldsymbol{\alpha}}$; we comment on this at the end of this subsection.
            
        In the limit $h\to\infty$, for a fixed root $\alpha$ and topological sector $\mathfrak{m}$ there is only a single slice $\mathcal{B}_{t,\boldsymbol{\alpha}}$ satisfying \eqref{eq.projection} which corresponds to the holomorphic sections of charge 
        \begin{equation}\label{eq.projectionWCP}
            t=\alpha(\mathfrak{m}).
        \end{equation}
        Moreover, it can be shown (in analogy to \cite{Lundin:2023tzw}) from dimensional reduction of the fields that the flat connections in the locus of the theory on $L^{2r-1}_{\boldsymbol{\alpha}}(h,\pm1)$, in the limit $h\to\infty$ give rise to gauge connections with flux on the two-cycle in $B^{r-1}_{\boldsymbol{\alpha}}$. This flux is of course labelled by $\mathfrak{m}$ whose components are now integral (e.g. for $G=U(N)$ we have $\mathfrak{m}=\diag(m_1,\dots,m_N)\in\mathbb{Z}^{N\times N}$). Hence, this method of reduction not only produces the one-loop determinant for the topologically trivial sector of the $(2r-2)$-dimensional theory but in fact for all possible topological sectors (cf. Kaluza-Klein reduction).
        
        The partition function on $B^{r-1}_{\boldsymbol{\alpha}}$ is then obtained by
        \begin{equation}
            \mathcal{Z}_{B^{r-1}_{\boldsymbol{\alpha}}}=\sum_{\mathfrak{m}}\int_{\mathfrak{h}}\dd a~\e{-S^\text{cl}_{2r-2}}\det{}^\prime_\text{adj}~S_r(\ii\alpha(a)|\boldsymbol{\R})|_{t=\alpha(\mathfrak{m})}.
        \end{equation}
        We note at this point that many previous results for the topologically twisted theory on closed four-manifolds are expressed in terms of a sum over equivariant fluxes on which certain stability conditions are imposed (e.g., \cite{Nekrasov:2003vi,Bershtein:2015xfa,Bershtein:2016mxz,Hosseini:2018uzp,Crichigno:2018adf,Bonelli:2020xps}). Here instead, we sum over the ``physical'' flux on the two-cycle in $B^{r-1}_{\boldsymbol{\alpha}}$.
        Explicit expressions for the one-loop determinants in each flux sector and classical actions $S^\text{cl}_{2r-1}$ on $B^{r-1}_{\boldsymbol{\alpha}}$ are presented in \autoref{sec--4} and \ref{sec--5}. Instanton contributions to $\mathcal{Z}_{B^{r-1}_{\boldsymbol{\alpha}}}$ will be discussed in \autoref{sec--7}. 
            
        \paragraph{Topologically Twisted and Exotic Theories.}\label{sec--3.3}
            Before we move on to explicit results, let us briefly comment on the different features of the topologically twisted and exotic theory whose perturbative partition function we obtained by reducing along the two different directions in \eqref{eq--2.free.vector.3d}, resp. \eqref{eq--2.free.vector.5d}. In 4d they are instances of a general cohomological twisting called \textit{Pestunization} \cite{Festuccia:2018rew,Festuccia:2019akm}. The topological twist is simply equivariant Donaldson-Witten theory. The canonical (and first) example of an exotic theory is Pestun's theory on $S^4$ \cite{Pestun:2007rz}.
            
            In cohomologically twisted theories the one-loop determinant is commonly computed from the equivariant index of a complex. Depending on whether we have a topological or an exotic theory, this complex will be either elliptic or transversally elliptic. This is reflected by the slices $\mathcal{B}^{t,\boldsymbol{\alpha}}$ being either compact or non-compact.
            
            In 4d, both for topological and exotic theories the partition function includes additional non-perturbative contributions from (anti-)instantons localised to the fixed points of a $U(1)$-isometry. While for the topological twist we have only instantons, the exotic theory allows for more generic distributions of instantons and anti-instantons at the fixed points. For instance, on $B^2_{\boldsymbol{\alpha}}$ we have anti-instantons at one of the three fixed points.

\section{Partition Function on \texorpdfstring{$S^3_{\boldsymbol{\alpha}}/S^1$}{S³ₐ/S¹}}\label{sec--4}

    We now set out to compute the integrand of $\mathcal{Z}_{B^1_{\boldsymbol{\alpha}}}$ explicitly. Before considering the reduction, however, we first perform some simplifications that arise from the specific form of the double sine function.   
    
    \subsection{Cancellations on \texorpdfstring{$S^3_{\boldsymbol{\alpha}}$}{S³ₐ}}

    We start computing the integrand of $\mathcal{Z}_{S^3_{\boldsymbol{\alpha}}}$ explicitly. Expressions will be different for the topological and exotic theories which we signify by a superscript. Before we implement the reduction procedure, on $S^3_{\boldsymbol{\alpha}}$, note that cancellations occur in 
    \begin{equation}
        Z_{S^3_{\boldsymbol{\alpha}}}(a,\boldsymbol{\R}):=\det{}^\prime_\text{adj}S_2(\ii\alpha(a)|\boldsymbol{\R})
    \end{equation} 
    which we take care of first. We start by rewriting the product over roots from the determinant as a product only over the positive roots:
    \begin{equation}\label{eq.ttop.perturbative.Salpha.1}
        Z^{\text{top}}_{S^{3}_{\boldsymbol{\alpha}}}=\prod_{\alpha\geq 0}\prod_{t\geq 0}\frac{\prod\limits_{n_1\in\mathcal{B}_{ t,\boldsymbol{\alpha}}}\left(\left(1-\frac{ \epsilon_1}{2}\right)\frac{t}{\alpha_1\alpha_2}+\frac{n_1}{\alpha_1}\epsilon_1+ \ii\alpha(a)\right)}{\prod\limits_{n_1\in\mathring{\mathcal{B}}_{ t,\boldsymbol{\alpha}}}\left(\left(1-\frac{ \epsilon_1}{2}\right)\frac{t}{\alpha_1\alpha_2}+\frac{n_1}{\alpha_1}\epsilon_1+ \ii\alpha(a)\right)}\prod_{t'\leq 0}\frac{\prod\limits_{n_1\in\mathcal{B}_{ t',\boldsymbol{\alpha}}}\left(\left(1-\frac{ \epsilon_1}{2}\right)\frac{t'}{\alpha_1\alpha_2}-\frac{n_1}{\alpha_1}\epsilon_1+ \ii\alpha(a)\right)}{\prod\limits_{n_1\in\mathring{\mathcal{B}}_{ t',\boldsymbol{\alpha}}}\left(\left(1-\frac{ \epsilon_1}{2}\right)\frac{t'}{\alpha_1\alpha_2}-\frac{n_1}{\alpha_1}\epsilon_1+ \ii\alpha(a)\right)},
    \end{equation}
    \begin{equation}\label{eq.tex.perturbative.Salpha.1}
        Z^{\text{ex}}_{S^3_{\boldsymbol{\alpha}}}=\prod_{\alpha\geq 0}\prod_{t\in\mathbb{Z}}\frac{\prod\limits_{n_1\in\mathcal{B}_{ t,\boldsymbol{\alpha}}}\left(-\frac{ \epsilon_1}{2}\frac{t}{\alpha_1\alpha_2}+\frac{n_1}{\alpha_1}\epsilon_1+ \ii\alpha(a)\right)}{\prod\limits_{n_1\in\mathring{\mathcal{B}}_{ t,\boldsymbol{\alpha}}}\left(-\frac{ \epsilon_1}{2}\frac{t}{\alpha_1\alpha_2}+\frac{n_1}{\alpha_1}\epsilon_1+ \ii\alpha(a)\right)}\prod_{t'\in\mathbb{Z}}\frac{\prod\limits_{n_1\in\mathcal{B}_{ t',\boldsymbol{\alpha}}}\left(-\frac{ \epsilon_1}{2}\frac{t'}{\alpha_1\alpha_2}-\frac{n_1}{\alpha_1}\epsilon_1+ \ii\alpha(a)\right)}{\prod\limits_{n_1\in\mathring{\mathcal{B}}_{ t',\boldsymbol{\alpha}}}\left(-\frac{ \epsilon_1}{2}\frac{t'}{\alpha_1\alpha_2}-\frac{n_1}{\alpha_1}\epsilon_1+ \ii\alpha(a)\right)},
    \end{equation}
    where we used \eqref{eq.epsilon.parameters} and \eqref{eq.squashingcontraint} to introduce the equivariance parameter $\epsilon_1$. Note that one product in $Z^\text{top}_{\boldsymbol{\alpha}}$ is over $t\ge0$ and one over $t'\le0$. Let us also introduce $a_\pm\in\mathbb{Z}$ satisfying
    \begin{equation}
        a_-\alpha_1-a_+\alpha_2=1,
    \end{equation}
    which always exist (by virtue of Bezout's lemma). In particular,
    \begin{equation}\label{eq.condition.apm}
        \gcd(a_-,\alpha_2)=1=\gcd(a_+,\alpha_1).
    \end{equation}
    In the following, we treat the topological and exotic case separately.

\begin{figure}[h!]
            \centering
            \begin{tikzpicture}
                \def\rad{.03cm};
                \draw[thick,-stealth] (-0.3,0)--(3.3,0) node [below]{$n_1$};
                \draw[thick,-stealth] (0,-0.3)--(0,3.3) node [left]{$n_2$};
                \draw[thick,-stealth] (0,0)--(3.1,3.1) node [above]{$\Vec{\X}^\text{top}$};
                \draw[thin] (0.5,0.5)--(3.2,0.5); 
                \draw[thin] (0.5,0.5)--(0.5,3.2);
                \draw[dashed] (1.5,0)--(0,1.5) node [left]{$t=3$};
                \draw[dashed] (2.5,0)--(0,2.5) node [left]{$t=5$};
                \foreach \x in {.5,1,...,3}{
                    \foreach \y in {.5,1,...,3}{
                        \draw[fill=black] (\x,\y) circle [radius=\rad];
                    }
                };
                \foreach \n in {.5,1,...,3}{
                    \draw[fill=light] (0,\n) circle [radius=2*\rad];
                    \draw[fill=light] (\n,0) circle [radius=2*\rad];
                }
                \draw[fill=light] (0,0) circle [radius=2*\rad];
            \end{tikzpicture}
            \hspace{0em}
            \begin{tikzpicture}
                \def\rad{.03cm};
                \draw[thick,-stealth] (-0.3,0)--(3.3,0) node [below]{$n_1$};
                \draw[thick,-stealth] (0,-0.3)--(0,3.3) node [left]{$n_2$};
                \draw[thick,-stealth] (0,0)--(2.2,3.3) node [above]{$\Vec{\X}^\text{top}$};
                \draw[thin] (0.5,0.5)--(3.2,0.5); 
                \draw[thin] (0.5,0.5)--(0.5,3.2);
                \draw[dashed] (3/4,0)--(0,0.5) node [left]{$t=3$};
                \draw[dashed] (9/4,0)--(0,1.5) node [left]{$t=9$};
                \draw[dashed] (12/4,0)--(0,2) node [left]{$t=12$};
                \foreach \x in {.5,1,...,3}{
                    \foreach \y in {.5,1,...,3}{
                        \draw[fill=black] (\x,\y) circle [radius=\rad];
                    }
                };
                \foreach \n in {.5,1,...,3}{
                    \draw[fill=light] (0,\n) circle [radius=2*\rad];
                    \draw[fill=light] (\n,0) circle [radius=2*\rad];
                }
                \draw[fill=light] (0,0) circle [radius=2*\rad];
            \end{tikzpicture}
            \hspace{0em}
            \begin{tikzpicture}
                \def\rad{.03cm};
                \draw[thick,-stealth] (-0.3,0)--(3.3,0) node [below]{$n_1$};
                \draw[thick,-stealth] (0,-0.3)--(0,3.3) node [left]{$n_2$};
                \draw[thick,-stealth] (0,0)--(3/5*3.3,3.3) node [above]{$\Vec{\X}^\text{top}$};
                \draw[thin] (0.5,0.5)--(3.2,0.5); 
                \draw[thin] (0.5,0.5)--(0.5,3.2);
                \draw[dashed] (2.5/3,0)--(0,2.5/5) node [left]{$t=5$};
                \draw[dashed] (6/3,0)--(0,6/5) node [left]{$t=6$};
                \draw[dashed] (10/3,0)--(0,10/5) node [left]{$t=20$};
                \foreach \x in {.5,1,...,3}{
                    \foreach \y in {.5,1,...,3}{
                        \draw[fill=black] (\x,\y) circle [radius=\rad];
                    }
                };
                \foreach \n in {.5,1,...,3}{
                    \draw[fill=light] (0,\n) circle [radius=2*\rad];
                    \draw[fill=light] (\n,0) circle [radius=2*\rad];
                }
                \draw[fill=light] (0,0) circle [radius=2*\rad];
            \end{tikzpicture}
            \caption{We plot the cone for $S^3_{\boldsymbol{\alpha}}$ which is simply the lattice spanned by integers $n_1,n_2\geq 0$. For various values of $\alpha_1,\alpha_2$ we plot the regions $\mathcal{B}_{ t,\boldsymbol{\alpha}}$ and $\mathring{\mathcal{B}}_{ t,\boldsymbol{\alpha}}$ as dashed lines. These two regions appear, respectively, at the numerator and the denominator in the perturbative partition function. Thus, after cancellations, only light blue modes survive. These contribute at different values of $t^{\text{top}}$ depending on the values of $\alpha_1,\alpha_2$. From left to right: (i) $\alpha_1=\alpha_2=1$, (ii) $\alpha_1=3,\alpha_2=2$, (iii) $\alpha_1=5,\alpha_2=3$.}
            \label{figure1}
        \end{figure}
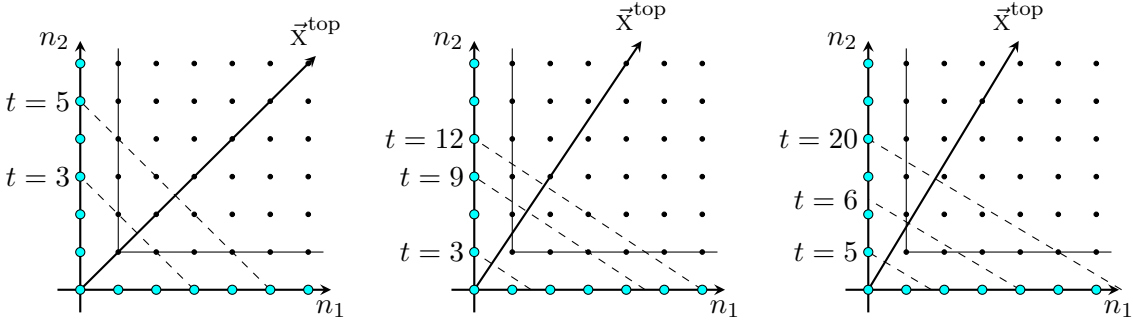

    \paragraph{Topological Theory.}
        Remember that the charge for rotations along the locally free $S^1$ is $t^\text{top}=\alpha_2n_1+\alpha_1n_2$. In order to identify the cancellations occurring it is useful to plot the slices $\mathcal{B}_{ t,\boldsymbol{\alpha}}$ and $\mathring{\mathcal{B}}_{ t,\boldsymbol{\alpha}}$ of the cone $\mathbb{Z}^2_{\ge0}$, see \autoref{figure1}. We notice that simplifications occur for points in the interior of the cone. Unpaired points are the ones on the $n_1$-and $n_2$-axis. Note that, while for $\alpha_1,\alpha_2=1$ two points survive the cancellations in each slice, for $\alpha_1\neq\alpha_2$ there might only be one or no such point at all.
        More explicitly, one can check that the bounds on $\tilde{n}_1=\frac{n_1}{\alpha_1}\in\mathbb{Z}_{\ge0}$ are the following:
        \begin{equation}\label{eq.bound.tpositive}
            t\geq0:\quad\begin{cases}\mathcal{B}_{ t,\boldsymbol{\alpha}}:&\rmd{a_+t}{\alpha_1}/\alpha_1\leq \tilde{n}_1\leq \frac{t}{\alpha_1\alpha_2}-\rmd{-a_-t}{\alpha_2}/\alpha_2\\
            \mathring{\mathcal{B}}_{ t,\boldsymbol{\alpha}}:&1-\rmd{-a_+t}{\alpha_1}/\alpha_1\leq \tilde{n}_1\leq \frac{t}{\alpha_1\alpha_2}-1+\rmd{a_-t}{\alpha_2}/\alpha_2,
        \end{cases}\end{equation}
        where $\rmd{x}{y}$ is the remainder of $x/y$. Similarly, the bounds for negative $t'$ are the following:
        \begin{equation}\label{eq.bound.tnegative}
            t'<0:\quad \begin{cases}\mathcal{B}_{ t',\boldsymbol{\alpha}}:&\frac{t'}{\alpha_1\alpha_2}+\rmd{a_-t'}{\alpha_2}/\alpha_2\leq \tilde{n}_1\leq-\rmd{-a_+t'}{\alpha_1}/\alpha_1\\
            \mathring{\mathcal{B}}_{ t',\boldsymbol{\alpha}}:&\frac{t'}{\alpha_1\alpha_2}+1-\rmd{-a_-t'}{\alpha_2}/\alpha_2\leq \tilde{n}_1\leq-1+\rmd{a_+t'}{\alpha_1}/\alpha_1.
        \end{cases}\end{equation}
        For $\alpha_1,\alpha_2=1$ these reduce to bounds appearing in \cite{Lundin:2021zeb} for the reduction from $S^3$ to $S^2$. In order to see how cancellations work in the general case, note that
        \begin{equation}\label{eq.relation.mmod}
            \rmd{a_+t}{\alpha_1}\neq 0~\Longrightarrow~1-\rmd{-a_+t}{\alpha_1}/\alpha_1=\rmd{a_+t}{\alpha_1}/\alpha_1,
        \end{equation}
        and similarly for $\rmd{a_-t}{\alpha_2}\neq0$. Therefore, depending on whether $\rmd{a_+t}{\alpha_1}=0$ and/or $\rmd{a_-t}{\alpha_2}=0$, only the term for $\tilde{n}_1=0$ and/or $\tilde{n}_1=t/\alpha_1\alpha_2$ survives. Thus, employing the condition in \eqref{eq.condition.apm}, we find\footnote{We henceforth combine $t\ge0$ and $t'\le0$ into a single $t\in\mathbb{Z}$.}
        \begin{equation}\begin{split}
            t\in\alpha_1\mathbb{Z}:\quad&\left(\left(1-\frac{ \epsilon_1}{2}\right)\frac{t}{\alpha_1\alpha_2}+\ii\alpha(a)\right),\\
            t\in\alpha_2\mathbb{Z}:\quad&\left(\left(1+\frac{ \epsilon_1}{2}\right)\frac{t}{\alpha_1\alpha_2}+\ii\alpha(a)\right).
        \end{split}\end{equation}
        We can now substitute in \eqref{eq.ttop.perturbative.Salpha.1}:  
        \begin{equation}\label{eq.top.simplified}
            Z^{\text{top}}_{S^3_{\boldsymbol{\alpha}}}=\prod_{\alpha>0}\prod_{t\in\alpha_1\mathbb{Z}}\left(\left(1-\frac{ \epsilon_1}{2}\right)\frac{t}{\alpha_1\alpha_2}+\ii\alpha(a)\right)\prod_{t\in\alpha_2\mathbb{Z}}\left(\left(1+\frac{ \epsilon_1}{2}\right)\frac{t}{\alpha_1\alpha_2}+\ii\alpha(a)\right).
        \end{equation}
        For $\alpha_1,\alpha_2=1$ this expression reproduces the perturbative partition function for $S^3$. For generic values of $\alpha_1,\alpha_2$ the slicing along $t^{\text{top}}$ only allows functions whose charge $t$ is a multiple of $\alpha_1$ or $\alpha_2$.

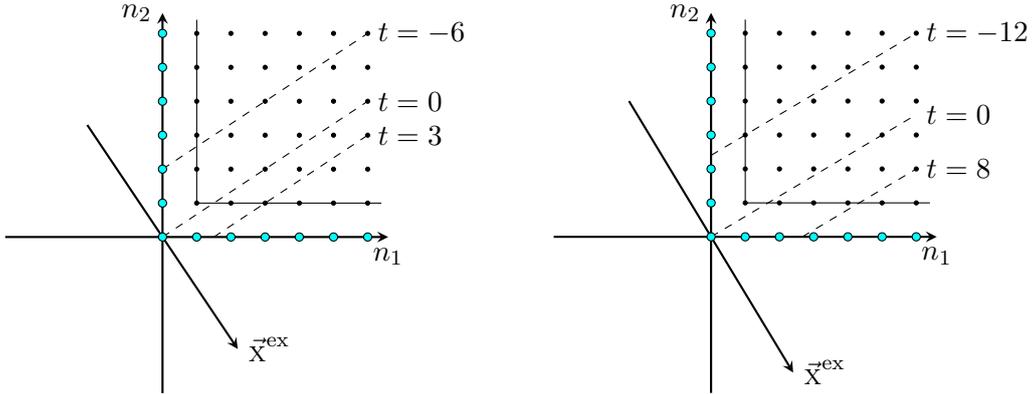
\begin{figure}[h!]
                \centering
                \begin{tikzpicture}[scale=.9]
                    \def\rad{.03cm};
                    \draw[thick,-stealth] (-2.3,0)--(3.3,0) node [below]{$n_1$};
                    \draw[thick,-stealth] (0,-2.3)--(0,3.3) node [left]{$n_2$};
                    \draw[thick,-stealth] (-2.2/2,3.3/2)--(2.2/2,-3.3/2) node [right]{$\Vec{\X}^\text{ex}$};
                    \draw[thin] (0.5,0.5)--(3.2,0.5); 
                    \draw[thin] (0.5,0.5)--(0.5,3.2);
                    \draw[dashed] (0,0)--(3,2) node [right]{$t=0$};
                    \draw[dashed] (3/4,0)--(3,1.5) node [right]{$t=3$};
                    \draw[dashed] (0,1)--(3,3) node [right]{$t=-6$};
                    \foreach \x in {.5,1,...,3}{
                        \foreach \y in {.5,1,...,3}{
                            \draw[fill=black] (\x,\y) circle [radius=\rad];
                        }
                    };
                    \foreach \n in {.5,1,...,3}{
                        \draw[fill=light] (0,\n) circle [radius=2*\rad];
                        \draw[fill=light] (\n,0) circle [radius=2*\rad];
                    }
                    \draw[fill=light] (0,0) circle [radius=2*\rad];
                \end{tikzpicture}
                \hspace{2em}
                \begin{tikzpicture}[scale=.9]
                    \def\rad{.03cm};
                    \draw[thick,-stealth] (-2.3,0)--(3.3,0) node [below]{$n_1$};
                    \draw[thick,-stealth] (0,-2.3)--(0,3.3) node [left]{$n_2$};
                    \draw[thick,-stealth] (-0.4*3,0.4*5)--(0.4*3,-0.4*5) node [right]{$\Vec{\X}^\text{ex}$};
                    \draw[thin] (0.5,0.5)--(3.2,0.5); 
                    \draw[thin] (0.5,0.5)--(0.5,3.2);
                    \draw[dashed] (0,0)--(3.,3.*3/5) node [right]{$t=0$};
                    \draw[dashed] (8/6,0)--(3,1) node [right]{$t=8$};
                    \draw[dashed] (0,12/10)--(3,3) node [right]{$t=-12$};
                    \foreach \x in {.5,1,...,3}{
                        \foreach \y in {.5,1,...,3}{
                            \draw[fill=black] (\x,\y) circle [radius=\rad];
                        }
                    };
                    \foreach \n in {.5,1,...,3}{
                        \draw[fill=light] (0,\n) circle [radius=2*\rad];
                        \draw[fill=light] (\n,0) circle [radius=2*\rad];
                    }
                    \draw[fill=light] (0,0) circle [radius=2*\rad];
                \end{tikzpicture}
                \caption{We plot the cone for $S^3_{\boldsymbol{\alpha}}$ which is the lattice spanned by integers $n_1,n_2\geq 0$. For various values of $\alpha_1,\alpha_2$ we plot the regions $\mathcal{B}_{ t,\boldsymbol{\alpha}}$ and $\mathring{\mathcal{B}}_{ t,\boldsymbol{\alpha}}$ as dashed lines. These two regions appear, respectively, at the numerator and the denominator in the perturbative partition function. Thus, after cancellations, only light blue modes survive. These contribute at different values of $t^{\text{ex}}$ depending on the values of $\alpha_1,\alpha_2$. From left to right: (i) $\alpha_1=3,\alpha_2=2$, (ii) $\alpha_1=5,\alpha_2=3$.}
                \label{figure2}
        \end{figure}

    \paragraph{Exotic Theory.}
        We follow a similar procedure here, where $t^{\text{ex}}=\alpha_2n_1-\alpha_1n_2$. Plots of the slices $\mathcal{B}_{ t,\boldsymbol{\alpha}},\mathring{\mathcal{B}}_{ t,\boldsymbol{\alpha}}$ of $\mathbb{Z}^2_{\ge0}$ are shown in \autoref{figure2}. Again, only modes on the $n_1,n_2$-axis survive after cancellations.
        Explicitly, the bounds on $\tilde{n}_1=\frac{n_1}{\alpha_1}\in\mathbb{Z}_{\geq 0}$ are the following:
        \begin{align}
            t\ge0:&~\begin{cases}
            \mathcal{B}_{ t,\boldsymbol{\alpha}}:&\tilde{n}_1\geq\rmd{a_+t}{\alpha_1}/\alpha_2,\\
            \mathring{\mathcal{B}}_{ t,\boldsymbol{\alpha}}:&\tilde{n}_1\geq1-\rmd{-a_+t}{\alpha_1}/\alpha_2,
            \end{cases}\\
            t'<0:&~\begin{cases}
                \mathcal{B}_{ t,\boldsymbol{\alpha}}:&\tilde{n}_1\geq-\frac{t}{\alpha_1\alpha_2}+\rmd{-a_-t}{\alpha_1}/\alpha_2,\\
                \mathring{\mathcal{B}}_{ t,\boldsymbol{\alpha}}:&\tilde{n}_1\geq-\frac{t}{\alpha_1\alpha_2}+1-\rmd{a_-t}{\alpha_2}/\alpha_2.
            \end{cases}
        \end{align}
        Similarly to the previous case, due to \eqref{eq.relation.mmod}, a mode survives cancellations if either $\rmd{a_+t}{\alpha_1}=0$ or $\rmd{a_-t}{\alpha_2}=0$. The same bounds hold for $t'=-t$.
        Combining the contributions from both $t$ and $t'$ in \eqref{eq.tex.perturbative.Salpha.1} and employing \eqref{eq.condition.apm}, we find
        \begin{equation}\label{eq.ex.simplified}
            Z^{\text{ex}}_{S^3_{\boldsymbol{\alpha}}}=\prod_{\alpha>0}\prod_{t\in\alpha_1\mathbb{Z}}\left(-\frac{ \epsilon_1}{2}\frac{t}{\alpha_1\alpha_2}+\ii\alpha(a)\right)\prod_{t\in\alpha_2\mathbb{Z}}\left(+\frac{ \epsilon_1}{2}\frac{t}{\alpha_1\alpha_2}+\ii\alpha(a)\right).
        \end{equation}
        Again, for $\alpha_1,\alpha_2=1$ we reproduce the result for a smooth $S^3$. The difference to the topological twist, besides the shift in $t$, are the different relations between the squashing parameters and the equivariance parameters \eqref{eq.epsilon.parameters}.

    \subsection{Finite Quotient and Reduction}
        
        In \autoref{sec--3} we explained that the partition function on the branched lens space $L^3_{\boldsymbol{\alpha}}(h,\pm1)$ can be obtained simply by imposing the condition \eqref{eq.projection} on $t$. Doing so for $Z^\text{top}_{S^3_{\boldsymbol{\alpha}}}$ and $Z^\text{ex}_{S^3_{\boldsymbol{\alpha}}}$ at fixed $\mathfrak{m}$ gives\footnote{See \cite{Inglese:2023tyc} for related work.}
        \begin{equation}\label{eq.L.top.simplified}
            Z^{\text{top}}_{L^3_{\boldsymbol{\alpha}}(h,1)}=\prod_{\alpha>0}\prod_{\substack{t=\alpha(\mathfrak{m})\mmod h \\ t\in\alpha_1\mathbb{Z}}}\left(\left(1-\frac{ \epsilon_1}{2}\right)\frac{t}{\alpha_1\alpha_2}+\ii\alpha(a)\right)\prod_{\substack{t=\alpha(\mathfrak{m})\mmod h\\ t\in\alpha_2\mathbb{Z}}}\left(\left(1+\frac{ \epsilon_1}{2}\right)\frac{t}{\alpha_1\alpha_2}+\ii\alpha(a)\right),
        \end{equation}
        \begin{equation}\label{eq.L.ex.simplified}
            Z^{\text{ex}}_{L^3_{\boldsymbol{\alpha}}(h,-1)}=\prod_{\alpha>0}\prod_{\substack{t=\alpha(\mathfrak{m})\mmod h \\ t\in\alpha_1\mathbb{Z}}}\left(-\frac{ \epsilon_1}{2}\frac{t}{\alpha_1\alpha_2}+\ii\alpha(a)\right)\prod_{\substack{t=\alpha(\mathfrak{m})\mmod h \\ t\in\alpha_2\mathbb{Z}}}\left(+\frac{ \epsilon_1}{2}\frac{t}{\alpha_1\alpha_2}+\ii\alpha(a)\right).
        \end{equation}
        The partition function on the quotient space $B^1_{\boldsymbol{\alpha}}$ is obtained by taking the limit $h\to\infty$ as explained in \autoref{subsec--reduction}. This gives
        \begin{equation}\label{eq.1loop.spindle.top}
            Z^{\text{top}}_{B^1_{\boldsymbol{\alpha}}}=\prod_{\substack{\alpha>0\\\alpha(\mathfrak{m})\in\alpha_1\mathbb{Z}}}\left(\left(1-\frac{ \epsilon_1}{2}\right)\frac{\alpha(\mathfrak{m})}{\alpha_1\alpha_2}+\ii\alpha(a)\right)\prod_{\substack{\alpha>0\\\alpha(\mathfrak{m})\in\alpha_2\mathbb{Z}}}\left(\left(1+\frac{ \epsilon_1}{2}\right)\frac{\alpha(\mathfrak{m})}{\alpha_1\alpha_2}+\ii\alpha(a)\right),
        \end{equation}
        \begin{equation}\label{eq.1loop.spindle.ex}
            Z^{\text{ex}}_{B^1_{\boldsymbol{\alpha}}}=\prod_{\substack{\alpha>0\\\alpha(\mathfrak{m})\in\alpha_1\mathbb{Z}}}\left(-\frac{ \epsilon_1}{2}\frac{\alpha(\mathfrak{m})}{\alpha_1\alpha_2}+\ii\alpha(a)\right)\prod_{\substack{\alpha>0\\\alpha(\mathfrak{m})\in\alpha_1\mathbb{Z}}}\left(+\frac{ \epsilon_1}{2}\frac{\alpha(\mathfrak{m})}{\alpha_1\alpha_2}+\ii\alpha(a)\right).
        \end{equation}
        For $\alpha_1,\alpha_2=1$ and $\epsilon_1=1$ one recovers\footnote{To match with the expression on $S^2$ we need to perform a constant shift of the Coulomb branch parameter $a\rightarrow a-\ii\mathfrak{m}$. Such shift has also been discussed in \cite{Lundin:2021zeb,Lundin:2023tzw}. It will also appear in \autoref{app--A} in order to match with the result for the spindle in \cite{Inglese:2023wky}.} the result for a smooth $S^2$ \cite{Benini:2015noa,Benini:2012ui} while for $\alpha_1=\alpha_2$ we reproduce \cite{Mori:2015bro}. Note that $\epsilon_1$ is different in \eqref{eq.1loop.spindle.top} and \eqref{eq.1loop.spindle.ex} according to \eqref{eq.epsilon.parameters}.

        The classical action $S^\text{cl}_2$ on $B^1_{\boldsymbol{\alpha}}$ is obtained from the one on $L^3_{\boldsymbol{\alpha}}(h,\pm1)$ \eqref{eq.classicalL3} by taking the limit $h\to\infty$ such that the ratio $\ell:=k/h$ is kept fixed:
        \begin{equation}\label{eq.S2cl}
            S^\text{cl}_2=-\ii\pi\ell\rho_3\tr(a^2\pm\mathfrak{m}),
        \end{equation}
        where we have introduced the shorthand $\rho_3:=\frac{\alpha_1\alpha_2}{\omega_1\omega_2(\epsilon_1)}$.

        At the end of this section we want to point out an interesting observation. In \cite{Inglese:2023wky} the partition function on $\mathbb{CP}^1_{(N_1,N_2)}\times S^1$ has been computed for two different backgrounds, called twist and anti-twist (here $\mathbb{CP}^1_{(N_1,N_2)}$ denotes $(N_1,N_2)$-weighted projective space, also known as a spindle; see \autoref{app--geometry}). In \autoref{app--A} we compute its small radius-limit with respect to the $S^1$ and find that the resulting one-loop determinant on $\mathbb{CP}^1_{(N_1,N_2)}$ is in perfect agreement with \eqref{eq.1loop.spindle.top} and \eqref{eq.1loop.spindle.ex} for twist, resp. anti-twist under the identification $\alpha_i=N_i$, $i=1,2$. We will have more to say about this matching in \autoref{sec--7}.

\section{Partition Function on \texorpdfstring{$S^5_{\boldsymbol{\alpha}}/S^1$}{S⁵ₐ/S¹}}\label{sec--5}

    In this section we compute the partition function for an $\mathcal{N}=2$ vector multiplet, at each flux sector but at the trivial instanton sector, on spaces $B^2_{\boldsymbol{\alpha}}$ with surplus conical angles. Also in this case, we will show how topologically twisted and exotic theories are obtained from the same theory of an $\mathcal{N}=1$ vector multiplet on the branched cover of $S^5$. Unlike $r=2$, the perturbative partition function on  $S^5_{\boldsymbol{\alpha}}$ does not admit any cancellations, as both terms in the triple sine function appear at the numerator. This both simplifies the computations and gives a richer dependence on the parameters $\alpha_i$.   

    \subsection{Finite Quotient and Reduction}
    Contrary to the case of $r=2$, here both products in the triple sine function \eqref{eq.tperturbative.Salpha} are in the numerator and no cancellations occur. Thus, we can start with the reduction procedure right away. Similar to $r=2$ we denote the one-loop piece of the partition function by
    \begin{equation}\label{eq--5d.Salpha}
        Z_{S^5_{\boldsymbol{\alpha}}}(a,\boldsymbol{\R}):=\det{}^\prime_\text{adj}S_3(\ii\alpha(a)|\boldsymbol{\R})
    \end{equation} 
    and also introduce the special function \cite{Pestun:2007rz,Festuccia:2018rew,Hama:2012bg,Lundin:2023tzw}
    \begin{equation}\label{eq.upsilon}
        \Upsilon^{\mathcal{B}_{ t,\boldsymbol{\alpha}}}\left(z|x,y\right)=\prod_{(n_1,n_2)\in\mathcal{B}_{ t,\boldsymbol{\alpha}}}\left(xn_1+yn_2+z\right)\prod_{(n_1,n_2)\in\mathring{\mathcal{B}}_{ t,\boldsymbol{\alpha}}}\left(x n_1+y n_2+\Bar{z}\right),
    \end{equation}
    in terms of which we can write the triple sine function as
    \begin{equation}
        S_3(\ii\alpha(a)|\boldsymbol{\R})=\prod_t\Upsilon^{\mathcal{B}_{t,\boldsymbol{\alpha}}}\left(\ii\alpha(a) \pm\omega_3 \frac{t}{\lcm\boldsymbol{\alpha}}\bigg|\frac{\epsilon_1}{\alpha_1},\frac{\epsilon_2}{\alpha_2}\right).
    \end{equation}
    Here, $\omega_3$ is obtained from \eqref{eq.squashingcontraint} as 
    \begin{equation}\begin{split}
        \text{top:}&\quad\omega_3=+1-\frac{\epsilon_1+\epsilon_2}{3},\\
        \text{ex:}&\quad\omega_3=-\frac{1}{3}+\frac{\epsilon_1+\epsilon_2}{3}.
    \end{split}\end{equation}

        In \autoref{sec--3} we explained that the partition function on $L^5_{\boldsymbol{\alpha}}(h,\pm1)$ can be obtained from \eqref{eq--5d.Salpha} simply by imposing the condition \eqref{eq.projection} on $t$. For fixed $\mathfrak{m}$ this gives
        \begin{equation}\label{eq.partitionfunction.L5alpha.ups.top}
            Z^{\text{top}}_{L^5_{\boldsymbol{\alpha}}(h, 1)}= \prod_{\alpha\in\Delta}\prod_{t=\alpha(\mathfrak{m})\mmod h}\Upsilon^{\mathcal{B}_{ t,\boldsymbol{\alpha}}}\left(\ii\alpha(a)+\left(1-\frac{\epsilon_1+\epsilon_2}{3}\right)\frac{t}{\lcm\boldsymbol{\alpha}}\bigg|\frac{\epsilon_1}{\alpha_1},\frac{\epsilon_2}{\alpha_2}\right),
        \end{equation}
        \begin{equation}\label{eq.partitionfunction.L5alpha.ups.ex}
            Z^{\text{ex}}_{L^5_{\boldsymbol{\alpha}}(h,- 1)}= \prod_{\alpha\in\Delta}\prod_{t=\alpha(\mathfrak{m})\mmod h}\Upsilon^{\mathcal{B}_{ t,\boldsymbol{\alpha}}}\left(\ii\alpha(a)\left(\frac{1}{3}-\frac{\epsilon_1+\epsilon_2}{3}\right)\frac{t}{\lcm\boldsymbol{\alpha}}\bigg|\frac{\epsilon_1}{\alpha_1},\frac{\epsilon_2}{\alpha_2}\right),
        \end{equation}
        where $\Delta$ denotes the root set of $G$.
        The partition function on the quotient space $B^1_{\boldsymbol{\alpha}}$ is obtained by taking the limit $h\to\infty$ as explained in \autoref{subsec--reduction}. This gives
        \begin{equation}\label{eq.CP2alphatop}
            Z^{\text{top}}_{B^{2}_{\boldsymbol{\alpha}}}=\prod_{\alpha\in\Delta}\Upsilon^{\mathcal{B}_{ \mathfrak{m},\boldsymbol{\alpha}}}\left(\ii\alpha(a)+\left(1-\frac{\epsilon_1+\epsilon_2}{3}\right)\frac{\alpha(\mathfrak{m})}{\lcm\boldsymbol{\alpha}}\bigg|\frac{\epsilon_1}{\alpha_1},\frac{\epsilon_2}{\alpha_2}\right),
        \end{equation}
        \begin{equation}\label{eq.CP2alphaex}
            Z^{\text{ex}}_{B^{2}_{\boldsymbol{\alpha}}}=\prod_{\alpha\in\Delta}\Upsilon^{\mathcal{B}_{ \mathfrak{m},\boldsymbol{\alpha}}}\left(\ii\alpha(a)+\left(\frac{1}{3}-\frac{\epsilon_1+\epsilon_2}{3}\right)\frac{\alpha(\mathfrak{m})}{\lcm\boldsymbol{\alpha}}\bigg|\frac{\epsilon_1}{\alpha_1},\frac{\epsilon_2}{\alpha_2}\right).
        \end{equation}
        Note that the slices $\mathcal{B}_{t,\boldsymbol{\alpha}}$ in the exotic case are non-compact and the products infinite. Thus, in order to make sense of \eqref{eq.CP2alphaex} they need to be suitably regulated, e.g. using zeta-function regularisation.

        The classical action $S^\text{cl}_4$ on $B^2_{\boldsymbol{\alpha}}$ is obtained from the one on $L^5_{\boldsymbol{\alpha}}(h,\pm1)$ by taking the limit $h\to\infty$ such that the product $g^2_\text{YM,4d}:=g^2_\text{YM}\cdot h$ is kept fixed:
        \begin{equation}\label{eq.S4cl}
            S^\text{cl}_4=\frac{(2\pi)^3}{g^2_\text{YM,4d}}\rho_5\tr a^2,
        \end{equation}
        where we have introduced the shorthand $\rho_5=\frac{\alpha_1\alpha_2\alpha_3}{\omega_1\omega_2\omega_3(\epsilon_1,\epsilon_2)}$. Note that, contrary to the $r=2$ case, the classical action has no flux-dependence\footnote{As discussed in \cite{Lundin:2023tzw}, such dependence can be achieved by a constant shift of the Coulomb branch parameter proportional to $\mathfrak{m}$.}.

    \subsection{Examples}
    
        The essential information needed to compute \eqref{eq.CP2alphatop}, resp. \eqref{eq.CP2alphaex} for explicit examples are the slices $\mathcal{B}_{\mathfrak{m},\boldsymbol{\alpha}}$. In this subsection we consider two different types of slicings corresponding to choices where
        \begin{align}\label{choices}
            \text{(i)}&\quad\gcd(\alpha_i,\alpha_j)=1\nonumber\\
            \text{(ii)}&\quad\gcd(\alpha_1,\alpha_2)=k_1,~\gcd(\alpha_2,\alpha_3)=k_3,~\gcd(\alpha_1,\alpha_3)=k_2\nonumber
        \end{align}
        and illustrate them for specific examples\footnote{The constraint $\gcd(\alpha_1,\alpha_2,\alpha_3)=1$ implies $\gcd(k_i,k_j)=1$ for all $i,j=1,2,3$.}. Note that the slices will always be compact (resp. non-compact) for the topological (resp. exotic) twist.
        
\begin{figure}[h!]\centering\tdplotsetmaincoords{80}{45}
                \begin{tikzpicture}[scale=0.75,tdplot_main_coords]
                    \draw[thick,-stealth] (0,0,0) -- (5,0,0) node[anchor=west]{$n_1$};
                    \draw[thick,-stealth] (0,0,0) -- (0,5,0) node[anchor=west]{$n_2$};
                    \draw[thick,-stealth] (0,0,0) -- (0,0,5) node[anchor=south]{$n_3$};
                    \draw[thick,-stealth] (0,0,0) -- (6/1.5,3/1.5,2/1.5) node[anchor=south]{$\Vec{\X}^\text{top}$};
                    \filldraw[draw=gray,fill=gray!20,opacity=0.3]     
                        (0,0,0)
                        -- (5,0,0)
                        -- (0,5,0)
                        -- cycle;
                    \filldraw[draw=gray,fill=gray!20,opacity=0.3]     
                        (0,0,0)
                        -- (0,5,0)
                        -- (0,0,5)
                        -- cycle;
                    \filldraw[draw=gray,fill=gray!20,opacity=0.3]     
                        (0,0,0)
                        -- (5,0,0)
                        -- (0,0,5)
                        -- cycle;
                    \filldraw[thick,draw=blue,fill=blue!20,opacity=0.5]       
                        (5*1/6,0,0)
                        -- (0,5*1/3,0)
                        -- (0,0,5*1/2)
                        -- cycle;
                    \draw[thick,blue]    
                        (5*1/6,0,0)
                        -- (0,5*1/3,0)
                        -- (0,0,5*1/2)
                        -- cycle;
                    \filldraw[thick,draw=red,fill=red!20,opacity=0.5]       
                        (5*2/6,0,0)
                        -- (0,5*2/3,0)
                        -- (0,0,5*2/2)
                        -- cycle;
                    \draw[thick,red]        
                        (5*2/6,0,0)
                        -- (0,5*2/3,0)
                        -- (0,0,5*2/2)
                        -- cycle;
                \end{tikzpicture}\hspace{6em}
                \tdplotsetmaincoords{80}{45}\begin{tikzpicture}[scale=0.75,tdplot_main_coords]
                    \draw[thick,-stealth] (0,0,0) -- (5,0,0) node[anchor=north]{$n_1$};
                    \draw[thick,-stealth] (0,0,0) -- (0,5,0) node[anchor=south]{$n_2$};
                    \draw[thick,-stealth] (0,0,0) -- (0,0,5) node[anchor=south]{$n_3$};
                    \draw[thick,-stealth] (0,0,0) -- (6/3,3/3,-2/3) node[anchor=west]{$\Vec{\X}^\text{ex}$}; 
                    \filldraw[draw=gray,fill=gray!20,opacity=0.3]     
                        (0,0,0)
                        -- (5,0,0)
                        -- (0,5,0)
                        -- cycle;
                    \filldraw[draw=gray,fill=gray!20,opacity=0.3]     
                        (0,0,0)
                        -- (0,5,0)
                        -- (0,0,5)
                        -- cycle;
                    \filldraw[draw=gray,fill=gray!20,opacity=0.3]     
                        (0,0,0)
                        -- (5,0,0)
                        -- (0,0,5)
                        -- cycle;
                    \filldraw[draw=green,fill=green!20,opacity=0.5]       
                        (0,0,0)
                        -- (0,2,3)
                        -- (1,0,3)
                        -- cycle;
                    \draw[thick,green] (0,2,3) -- (0,0,0) -- (1,0,3);
                    \filldraw[draw=red,fill=red!20,opacity=0.5]       
                        (0,0,1)
                        -- (0,2,4)
                        -- (1,0,4)
                        -- cycle;
                    \draw[thick,red] (0,2,4)
                        -- (0,0,1)
                        -- (1,0,4);
                    \filldraw[draw=blue,fill=blue!20,opacity=0.5]       
                        (1,0,0)
                        -- (0,2,0)
                        -- (0,3+0.4,1.5+0.6)
                        -- (2-0.2,0,3-0.6)
                        -- cycle;
                    \draw[thick,blue] (2-0.2,0,3-0.6) -- (1,0,0) -- (0,2,0) -- (0,3+0.4,1.5+0.6);
                \end{tikzpicture}
                \caption{Cone for $S^5_{1,2,3}$. Left side: sliced along $\vec{\X}^\text{top}$ for $t=5$ (blue) and $t=10$ (red). At $t=0$ the slice only contains the origin. Right side: sliced along $\vec{\X}^\text{ex}$ for $t=6$ (blue), $t=0$ (green) and $t=-2$ (red). The slices are compact for $\vec{\X}^\text{top}$ and non-compact for $\vec{\X}^\text{ex}$.}
                \label{figure6b}
            \end{figure}
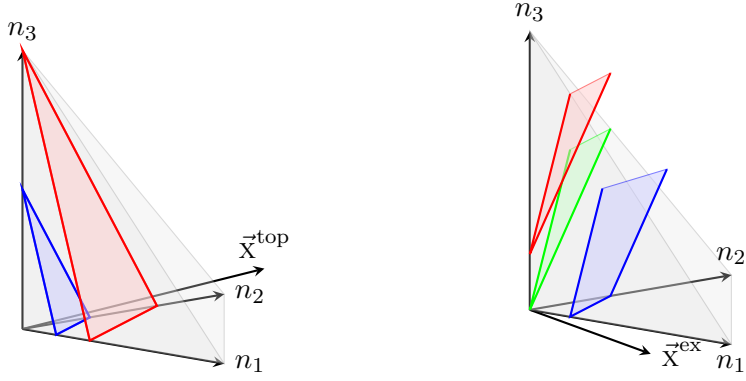

        \paragraph{Choice (i).}\label{bar}
            For definiteness let us start by looking at $\alpha_1=1,\alpha_2=2,\alpha_3=3$. Slices $\mathcal{B}_{t,\boldsymbol{\alpha}}\subset\mathbb{Z}^3_{\ge0}$ are depicted in \autoref{figure6b} for topological and exotic theories.
            For $h\to\infty$, only a single slice survives and contributes at the flux sector $t=\alpha(\mathfrak{m})$. We show the $(n_1,n_2)$-projection of such slices for the topological and exotic theories for two different choices of $\mathfrak{m}$ in \autoref{figure7b} and \ref{figure7c}.
            \begin{figure}[h!]\centering\tdplotsetmaincoords{0}{0}
                \begin{tikzpicture}[scale=0.75,tdplot_main_coords]
                    \draw[thick,-stealth] (-0.5,0,0) -- (5.2,0,0) node[anchor=north]{$n_1$};
                    \draw[thick,-stealth] (0,-0.5,0) -- (0,5.2,0) node[anchor=east]{$n_2$};
                    \draw[step=1.0,gray!60] (-.5,-.5) grid (5.2,5.2);
                    \filldraw[thick,draw=blue,fill=blue!20,opacity=0.5]       
                        (2,0,0)
                        -- (0,4,0)
                        -- (0,0,2)
                        -- cycle;
                    \draw[thick,blue]    
                        (2,0,0)
                        -- (0,4,0)
                        -- (0,0,2)
                        -- cycle;
                    \draw[fill=light] (0,0) circle [radius=0.1cm];
                    \draw[fill=light] (1,0) circle [radius=0.1cm];
                    \draw[fill=light] (2,0) circle [radius=0.1cm];
                    \draw[fill=light] (0,2) circle [radius=0.1cm];
                    \draw[fill=light] (0,2) circle [radius=0.1cm];
                    \draw[fill=light] (1,2) circle [radius=0.1cm];
                    \draw[fill=light] (0,4) circle [radius=0.1cm];
                \end{tikzpicture}\hspace{6em}
                \tdplotsetmaincoords{0}{0}
                \begin{tikzpicture}[scale=0.75,tdplot_main_coords]
                    \draw[thick,-stealth] (-0.5,0,0) -- (5.2,0,0) node[anchor=north]{$n_1$};
                    \draw[thick,-stealth] (0,-0.5,0) -- (0,5.2,0) node[anchor=east]{$n_2$};
                    \draw[step=1.0,gray!60] (-.5,-.5) grid (5.2,5.2); 
                    \filldraw[draw=blue,fill=blue!20,opacity=0.5]       
                        (2,0,0)
                        -- (5,0,0)
                        -- (5,5,0)
                        -- (0,5,1)
                        -- (0,4,1)
                        -- cycle;
                    \draw[thick,blue] (5,0,1) -- (2,0,0) -- (0,4,0) -- (0,5,1);
                    \draw[fill=light] (2,0) circle [radius=0.1cm];
                    \draw[fill=light] (3,0) circle [radius=0.1cm];
                    \draw[fill=light] (4,0) circle [radius=0.1cm];
                    \draw[fill=light] (5,0) circle [radius=0.1cm];
                    \draw[fill=light] (1,2) circle [radius=0.1cm];
                    \draw[fill=light] (2,2) circle [radius=0.1cm];
                    \draw[fill=light] (3,2) circle [radius=0.1cm];
                    \draw[fill=light] (4,2) circle [radius=0.1cm];
                    \draw[fill=light] (5,2) circle [radius=0.1cm];
                    \draw[fill=light] (0,4) circle [radius=0.1cm];
                    \draw[fill=light] (1,4) circle [radius=0.1cm];
                    \draw[fill=light] (2,4) circle [radius=0.1cm];
                    \draw[fill=light] (3,4) circle [radius=0.1cm];
                    \draw[fill=light] (4,4) circle [radius=0.1cm];
                    \draw[fill=light] (5,4) circle [radius=0.1cm];
                \end{tikzpicture}
                \caption{Slices $\mathcal{B}_{\mathfrak{m},\boldsymbol{\alpha}}$ for $B^2_{1,2,3}$ and $\alpha(\mathfrak{m})=12$. Left side: topologically twisted theory. Right side: Exotic theory. Only points in light blue contribute to the slices.}
                \label{figure7b}
            \end{figure}
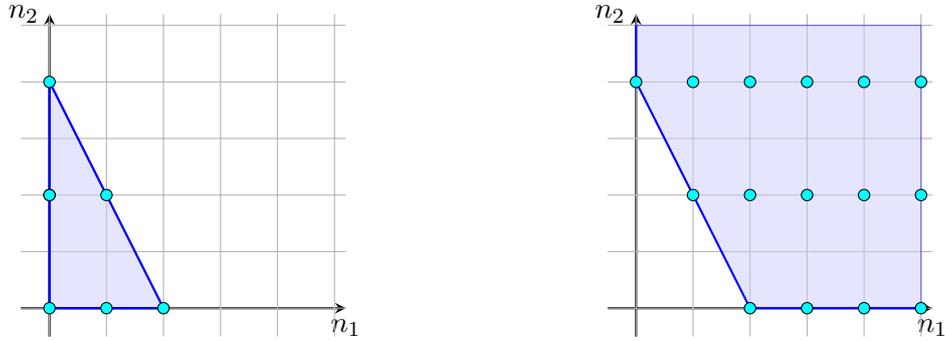
            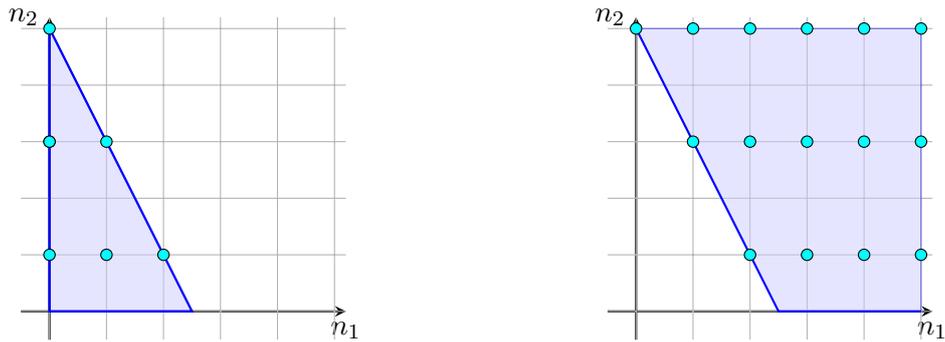
\begin{figure}[h!]\centering\tdplotsetmaincoords{0}{0}
                \begin{tikzpicture}[scale=0.75,tdplot_main_coords]
                    \draw[thick,-stealth] (-0.5,0,0) -- (5.2,0,0) node[anchor=north]{$n_1$};
                    \draw[thick,-stealth] (0,-0.5,0) -- (0,5.2,0) node[anchor=east]{$n_2$};
                    \draw[step=1.0,gray!60] (-.5,-.5) grid (5.2,5.2);
                    \filldraw[thick,draw=blue,fill=blue!20,opacity=0.5]       
                        (5/2,0,0)
                        -- (0,5,0)
                        -- (0,0,2)
                        -- cycle;
                    \draw[thick,blue]    
                        (5/2,0,0)
                        -- (0,5,0)
                        -- (0,0,2)
                        -- cycle;
                    \draw[fill=light] (0,1) circle [radius=0.1cm];
                    \draw[fill=light] (1,1) circle [radius=0.1cm];
                    \draw[fill=light] (2,1) circle [radius=0.1cm];
                    \draw[fill=light] (0,3) circle [radius=0.1cm];
                    \draw[fill=light] (0,3) circle [radius=0.1cm];
                    \draw[fill=light] (1,3) circle [radius=0.1cm];
                    \draw[fill=light] (0,5) circle [radius=0.1cm];
                \end{tikzpicture}\hspace{6em}
                \tdplotsetmaincoords{0}{0}
                \begin{tikzpicture}[scale=0.75,tdplot_main_coords]
                    \draw[thick,-stealth] (-0.5,0,0) -- (5.2,0,0) node[anchor=north]{$n_1$};
                    \draw[thick,-stealth] (0,-0.5,0) -- (0,5.2,0) node[anchor=east]{$n_2$};
                    \draw[step=1.0,gray!60] (-.5,-.5) grid (5.2,5.2); 
                    \filldraw[draw=blue,fill=blue!20,opacity=0.5]       
                        (5/2,0,0)
                        -- (5,0,0)
                        -- (5,5,0)
                        -- (0,5,1)
                        -- cycle;
                    \draw[thick,blue] (5,0,1) -- (5/2,0,0) -- (0,5,0);
                    \draw[fill=light] (2,1) circle [radius=0.1cm];
                    \draw[fill=light] (3,1) circle [radius=0.1cm];
                    \draw[fill=light] (4,1) circle [radius=0.1cm];
                    \draw[fill=light] (5,1) circle [radius=0.1cm];
                    \draw[fill=light] (1,3) circle [radius=0.1cm];
                    \draw[fill=light] (2,3) circle [radius=0.1cm];
                    \draw[fill=light] (3,3) circle [radius=0.1cm];
                    \draw[fill=light] (4,3) circle [radius=0.1cm];
                    \draw[fill=light] (5,3) circle [radius=0.1cm];
                    \draw[fill=light] (0,5) circle [radius=0.1cm];
                    \draw[fill=light] (1,5) circle [radius=0.1cm];
                    \draw[fill=light] (2,5) circle [radius=0.1cm];
                    \draw[fill=light] (3,5) circle [radius=0.1cm];
                    \draw[fill=light] (4,5) circle [radius=0.1cm];
                    \draw[fill=light] (5,5) circle [radius=0.1cm];
                \end{tikzpicture}
                \caption{Slices $\mathcal{B}_{\mathfrak{m},\boldsymbol{\alpha}}$ for $B^2_{1,2,3}$ and $\alpha(\mathfrak{m})=15$. Left side: topologically twisted theory. Right side: Exotic theory. Only points in light blue contribute to the slices.}
                \label{figure7c}
            \end{figure}
            Looking at these slices we notice that $n_2$ only comes in multiples\footnote{Note that a similar behaviour occurs for $r=2$ in \autoref{sec--4}.} of $\alpha_2=2$. For generic values of $\alpha_1,\alpha_2,\alpha_3$ one can check that $n_i$ comes in multiples of $\alpha_i$. Let us introduce $\tilde{n}_i:=\frac{n_i}{\alpha_i}\in\mathbb{Z}$ and rewrite the partition functions as follows:
			\begin{equation}\label{eq.CP2alphatop2}
                Z^{\text{top}}_{B^{2}_{\boldsymbol{\alpha}}}=\prod_{\alpha\in\Delta}\Upsilon^{\widetilde{\mathcal{B}}_{ \mathfrak{m},\boldsymbol{\alpha}}}\left(\ii\alpha(a)+\left(1-\frac{\epsilon_1+\epsilon_2}{3}\right)\frac{\alpha(\mathfrak{m})}{\lcm\boldsymbol{\alpha}}\bigg|\epsilon_1,\epsilon_2\right),
            \end{equation}    
            \begin{equation}\label{eq.CP2alphaex2}
                Z^{\text{ex}}_{B^{2}_{\boldsymbol{\alpha}}}=\prod_{\alpha\in\Delta}\Upsilon^{\widetilde{\mathcal{B}}_{ \mathfrak{m},\boldsymbol{\alpha}}}\left(\ii\alpha(a)+\left(\frac{1}{3}-\frac{\epsilon_1+\epsilon_2}{3}\right)\frac{\alpha(\mathfrak{m})}{\lcm\boldsymbol{\alpha}}\bigg|\epsilon_1,\epsilon_2\right),
            \end{equation} 
            where, the new slices are defined for $\rmd{\alpha(\mathfrak{m})}{\alpha_i}=0,\,i=1,2$, as
            \begin{equation}\begin{split}
                \text{top:}\quad&\widetilde{\mathcal{B}}_{\mathfrak{m},\boldsymbol{\alpha}}=\{(\tilde{n}_1,\tilde{n}_2)\in\mathbb{Z}^2_{\geq 0}\; |\; \tilde{n}_1+\tilde{n}_2\leq \alpha(\mathfrak{m})/(\alpha_1\alpha_2\alpha_3)\},\\
                \text{ex:}\quad&\widetilde{\mathcal{B}}_{\mathfrak{m},\boldsymbol{\alpha}}=\{(\tilde{n}_1,\tilde{n}_2)\in\mathbb{Z}^2_{\geq 0}\; |\; \tilde{n}_1+\tilde{n}_2\geq \alpha(\mathfrak{m})/(\alpha_1\alpha_2\alpha_3)\}.
            \end{split}\end{equation}
            The slices $\widetilde{\mathcal{B}}_{\mathfrak{m},\boldsymbol{\alpha}}$ are shown in \autoref{figure7d} for $\alpha(\mathfrak{m})=12$. Note that whenever no light blue points appear in the plots, then all points contribute to the slice.
            \begin{figure}[h!]\centering\tdplotsetmaincoords{0}{0}
                \begin{tikzpicture}[scale=0.75,tdplot_main_coords]
                    \draw[thick,-stealth] (-0.5,0,0) -- (5.2,0,0) node[anchor=north]{$n_1$};
                    \draw[thick,-stealth] (0,-0.5,0) -- (0,5.2,0) node[anchor=east]{$n_2$};
                    \draw[step=1.0,gray!60] (-.5,-.5) grid (5.2,5.2);
                    \filldraw[thick,draw=blue,fill=blue!20,opacity=0.5]       
                        (2,0,0)
                        -- (0,2,0)
                        -- (0,0,2)
                        -- cycle;
                    \draw[thick,blue]    
                        (2,0,0)
                        -- (0,2,0)
                        -- (0,0,2)
                        -- cycle;
                \end{tikzpicture}\hspace{6em}
                \tdplotsetmaincoords{0}{0}
                \begin{tikzpicture}[scale=0.75,tdplot_main_coords]
                    \draw[thick,-stealth] (-0.5,0,0) -- (5.2,0,0) node[anchor=north]{$n_1$};
                    \draw[thick,-stealth] (0,-0.5,0) -- (0,5.2,0) node[anchor=east]{$n_2$};
                    \draw[step=1.0,gray!60] (-.5,-.5) grid (5.2,5.2); 
                    \filldraw[draw=blue,fill=blue!20,opacity=0.5]       
                        (2,0,0)
                        -- (5,0,0)
                        -- (5,5,0)
                        -- (0,5,1)
                        -- (0,2,1)
                        -- cycle;
                    \draw[thick,blue] (5,0,1) -- (2,0,0) -- (0,2,0) -- (0,5,1);
                \end{tikzpicture}
                \caption{Slices $\widetilde{\mathcal{B}}_{\mathfrak{m},\boldsymbol{\alpha}}$ for $B^2_{1,2,3}$ and $\alpha(\mathfrak{m})=12$. Left side: topologically twisted theory. Right side: Exotic theory. All points contribute to the slice.}
                \label{figure7d}
            \end{figure}
            For generic values of $\alpha(\mathfrak{m})$ the slices do not necessarily start from the origin and then certain shifts depending on $\rmd{\alpha(\mathfrak{m})}{\alpha_1}$ and $\rmd{\alpha(\mathfrak{m})}{\alpha_2}$ need to be included in the definition of $\widetilde{\mathcal{B}}_{\mathfrak{m},\boldsymbol{\alpha}}$.

            In the case $\alpha_1,\alpha_2,\alpha_3=1$ the result for both topological and exotic theories agree with those on $\mathbb{CP}^2$ in \cite{Lundin:2021zeb}. More interestingly, $Z^\text{top}_{B^2_{\boldsymbol{\alpha}}}$ is in agreement with the equivariant index on weighted projective space $\mathbb{CP}^2_{(N_1,N_2,N_3)}$ in \cite{Martelli:2023oqk} under the identification 
            \begin{equation}\label{eq.CP2.weights}
                N_1=\alpha_2\alpha_3,\qquad N_2=\alpha_1\alpha_3,\qquad N_3=\alpha_1\alpha_2.
            \end{equation} 
            Specifically, for any value of $\alpha(\mathfrak{m})$ we can match\footnote{Note that in \cite{Martelli:2023oqk} the flux $\alpha(\mathfrak{m})=T(\Lambda_1,\Lambda_2,\Lambda_3)$ is expressed in terms of equivariant fluxes $\Lambda_1,\Lambda_2,\Lambda_3$.} both, the size and the shape of $\widetilde{\mathcal{B}}_{\mathfrak{m},\boldsymbol{\alpha}}$ with the slices of \cite{Martelli:2023oqk}. However, in \cite{Martelli:2023oqk} topological sectors are labelled by equivariant fluxes rather than the ``physical'' flux in our results which obscures a direct comparison of the final expressions.

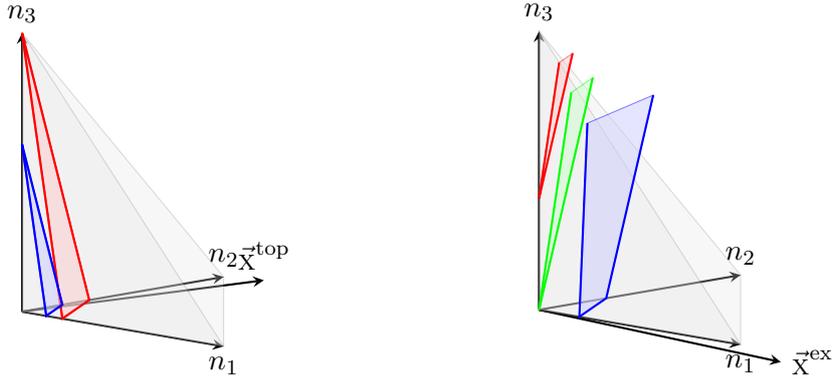
\begin{figure}[h!]\centering\tdplotsetmaincoords{80}{45}
                \begin{tikzpicture}[scale=0.75,tdplot_main_coords]
                    \draw[thick,-stealth] (0,0,0) -- (5,0,0) node[anchor=north]{$n_1$};
                    \draw[thick,-stealth] (0,0,0) -- (0,5,0) node[anchor=south]{$n_2$};
                    \draw[thick,-stealth] (0,0,0) -- (0,0,5) node[anchor=south]{$n_3$};
                    \draw[thick,-stealth] (0,0,0) -- (5*0.75,3*0.75,1*0.75) node[anchor=south]{$\Vec{\X}^\text{top}$};
                    \filldraw[draw=gray,fill=gray!20,opacity=0.3]     
                        (0,0,0)
                        -- (5,0,0)
                        -- (0,5,0)
                        -- cycle;
                    \filldraw[draw=gray,fill=gray!20,opacity=0.3]     
                        (0,0,0)
                        -- (0,5,0)
                        -- (0,0,5)
                        -- cycle;
                    \filldraw[draw=gray,fill=gray!20,opacity=0.3]     
                        (0,0,0)
                        -- (5,0,0)
                        -- (0,0,5)
                        -- cycle;
                    \filldraw[thick,draw=red,fill=red!20,opacity=0.5]       
                        (1,0,0)
                        -- (0,5/3,0)
                        -- (0,0,5)
                        -- cycle;
                    \draw[thick,red]    
                        (1,0,0)
                        -- (0,5/3,0)
                        -- (0,0,5)
                        -- cycle;
                    \filldraw[thick,draw=blue,fill=blue!20,opacity=0.5]       
                        (3/5,0,0)
                        -- (0,1,0)
                        -- (0,0,3)
                        -- cycle;
                    \draw[thick,blue]        
                        (3/5,0,0)
                        -- (0,1,0)
                        -- (0,0,3)
                        -- cycle;
                \end{tikzpicture}\hspace{6em}
                \tdplotsetmaincoords{80}{45}
                \begin{tikzpicture}[scale=0.75,tdplot_main_coords]
                    \draw[thick,-stealth] (0,0,0) -- (5,0,0) node[anchor=north]{$n_1$};
                    \draw[thick,-stealth] (0,0,0) -- (0,5,0) node[anchor=south]{$n_2$};
                    \draw[thick,-stealth] (0,0,0) -- (0,0,5) node[anchor=south]{$n_3$};
                    \draw[thick,-stealth] (0,0,0) -- (5*0.75,3*0.75,-1*0.75) node[anchor=west]{$\Vec{\X}^\text{ex}$}; 
                    \filldraw[draw=gray,fill=gray!20,opacity=0.3]     
                        (0,0,0)
                        -- (5,0,0)
                        -- (0,5,0)
                        -- cycle;
                    \filldraw[draw=gray,fill=gray!20,opacity=0.3]     
                        (0,0,0)
                        -- (0,5,0)
                        -- (0,0,5)
                        -- cycle;
                    \filldraw[draw=gray,fill=gray!20,opacity=0.3]     
                        (0,0,0)
                        -- (5,0,0)
                        -- (0,0,5)
                        -- cycle;
                    \filldraw[draw=green,fill=green!20,opacity=0.5]       
                        (0,0,0)
                        -- (0,1+1/3,3+1)
                        -- (1-1/5,0,5-1)
                        -- cycle;
                    \draw[thick,green] (0,1+1/3,3+1) -- (0,0,0) -- (1-1/5,0,5-1);
                    \filldraw[draw=red,fill=red!20,opacity=0.5]       
                        (0,0,2)
                        -- (0,1/3+1.5/3,3+1.5)
                        -- (1/5+1.5/5,0,3+1.5)
                        -- cycle;
                    \draw[thick,red] (0,1/3+1.5/3,3+1.5) -- (0,0,2) -- (1/5+1.5/5,0,3+1.5);
                    \filldraw[draw=blue,fill=blue!20,opacity=0.5]       
                        (1,0,0)
                        -- (0,5/3,0)
                        -- (0,5/2+1/3,5/2+1)
                        -- (1+1/5,0,5/2+1)
                        -- cycle;
                    \draw[thick,blue] (1+1/5,0,5/2+1) -- (1,0,0) -- (0,5/3,0) -- (0,5/2+1/3,5/2+1);
                \end{tikzpicture}
                \caption{Cone for $S^5_{3,5,15}$. Left side: sliced along $\vec{\X}^\text{top}$ for $t=3$ (blue) and $t=5$ (red). At $t=0$ the slice only contains the origin. Right side: sliced along $\vec{\X}^\text{ex}$ for $t=5$ (blue), $t=0$ (green) and $t=-2$ (red). The slices are compact for $\vec{\X}^\text{top}$ and non-compact for $\vec{\X}^\text{ex}$.}
                \label{figure8}
            \end{figure}

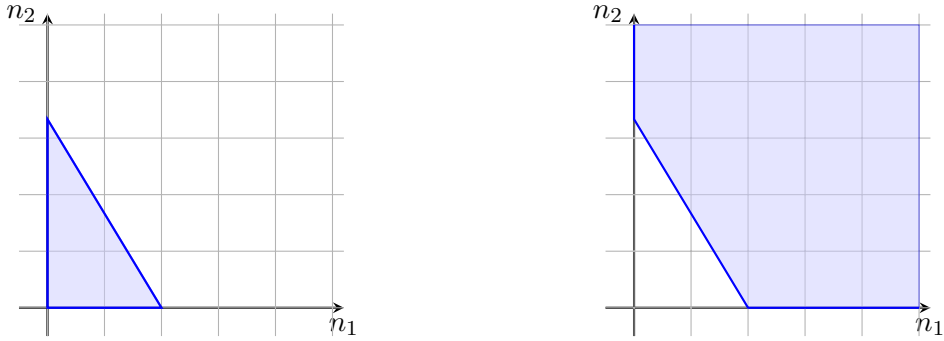
\begin{figure}[h!]\centering\tdplotsetmaincoords{0}{0}
                \begin{tikzpicture}[scale=0.75,tdplot_main_coords]
                    \draw[thick,-stealth] (-0.5,0,0) -- (5.2,0,0) node[anchor=north]{$n_1$};
                    \draw[thick,-stealth] (0,-0.5,0) -- (0,5.2,0) node[anchor=east]{$n_2$};
                    \draw[step=1.0,gray!60] (-.5,-.5) grid (5.2,5.2);
                    \filldraw[thick,draw=blue,fill=blue!20,opacity=0.5]       
                        (2,0,0)
                        -- (0,10/3,0)
                        -- (0,0,5)
                        -- cycle;
                    \draw[thick,blue]    
                        (2,0,0)
                        -- (0,10/3,0)
                        -- (0,0,5)
                        -- cycle;
                \end{tikzpicture}\hspace{6em}
                \tdplotsetmaincoords{0}{0}
                \begin{tikzpicture}[scale=0.75,tdplot_main_coords]
                    \draw[thick,-stealth] (-0.5,0,0) -- (5.2,0,0) node[anchor=north]{$n_1$};
                    \draw[thick,-stealth] (0,-0.5,0) -- (0,5.2,0) node[anchor=east]{$n_2$};
                    \draw[step=1.0,gray!60] (-.5,-.5) grid (5.2,5.2); 
                    \filldraw[draw=blue,fill=blue!20,opacity=0.5]       
                        (2,0,0)
                        -- (0,10/3,0)
                        -- (0,5,0)
                        -- (5,5,4)
                        -- (5,0,0)
                        -- (4*2/3,0,5*2/3)
                        -- cycle;
                    \draw[thick,blue] 
                        (5,0,0)
                        -- (2,0,0)
                        -- (0,10/3,0)
                        -- (0,5,0);
                \end{tikzpicture}
                \caption{Slices for $B^2_{3,5,15}$. Left side: $\mathcal{B}_{\mathfrak{m},\boldsymbol{\alpha}}$ of the topologically twisted theory for $\alpha(\mathfrak{m})=10$. For $\alpha(\mathfrak{m})=0$ only the origin contributes. Right side: $\mathcal{B}_{\mathfrak{m},\boldsymbol{\alpha}}$ of the exotic theory for $\alpha(\mathfrak{m})=10$. For $\alpha(\mathfrak{m})\leq 0$ the entire quadrant contributes. All points contribute to the slice.}
                \label{figure9}
            \end{figure}

\begin{figure}[h!]\centering\tdplotsetmaincoords{0}{0}
                \begin{tikzpicture}[scale=0.75,tdplot_main_coords]
                    \draw[thick,-stealth] (-0.5,0,0) -- (5.2,0,0) node[anchor=north]{$n_1$};
                    \draw[thick,-stealth] (0,-0.5,0) -- (0,5.2,0) node[anchor=east]{$n_2$};
                    \draw[step=1.0,gray!60] (-.5,-.5) grid (5.2,5.2);
                    \filldraw[thick,draw=blue,fill=blue!20,opacity=0.5]       
                        (3,0,0)
                        -- (0,3,0)
                        -- (0,0,0)
                        -- cycle;
                    \draw[thick,blue]    
                        (3,0,0)
                        -- (0,3,0)
                        -- (0,0,3)
                        -- cycle;
                    \draw[fill=light] (1,0) circle [radius=0.1cm];
                    \draw[fill=light] (3,0) circle [radius=0.1cm];
                    \draw[fill=light] (0,1) circle [radius=0.1cm];
                    \draw[fill=light] (0,3) circle [radius=0.1cm];
                    \draw[fill=light] (1,2) circle [radius=0.1cm];
                    \draw[fill=light] (2,1) circle [radius=0.1cm];
                \end{tikzpicture}\hspace{6em}
                \tdplotsetmaincoords{0}{0}
                \begin{tikzpicture}[scale=0.75,tdplot_main_coords]
                    \draw[thick,-stealth] (-0.5,0,0) -- (5.2,0,0) node[anchor=north]{$n_1$};
                    \draw[thick,-stealth] (0,-0.5,0) -- (0,5.2,0) node[anchor=east]{$n_2$};
                    \draw[step=1.0,gray!60] (-.5,-.5) grid (5.2,5.2); 
                    \filldraw[draw=blue,fill=blue!20,opacity=0.5]       
                        (3,0,0)
                        -- (0,3,0)
                        -- (0,5,0)
                        -- (5,5,0)
                        -- (5,0,0)
                        -- cycle;
                    \draw[thick,blue] 
                        (5,0,0)
                        -- (3,0,0)
                        -- (0,3,0)
                        -- (0,5,0);
                    \draw[fill=light] (3,0) circle [radius=0.1cm];
                    \draw[fill=light] (0,3) circle [radius=0.1cm];
                    \draw[fill=light] (1,2) circle [radius=0.1cm];
                    \draw[fill=light] (2,1) circle [radius=0.1cm];
                    \draw[fill=light] (5,0) circle [radius=0.1cm];
                    \draw[fill=light] (0,5) circle [radius=0.1cm];
                    \draw[fill=light] (3,2) circle [radius=0.1cm];
                    \draw[fill=light] (2,3) circle [radius=0.1cm];
                    \draw[fill=light] (1,4) circle [radius=0.1cm];
                    \draw[fill=light] (4,1) circle [radius=0.1cm];
                    \draw[fill=light] (5,2) circle [radius=0.1cm];
                    \draw[fill=light] (2,5) circle [radius=0.1cm];
                    \draw[fill=light] (5,4) circle [radius=0.1cm];
                    \draw[fill=light] (4,5) circle [radius=0.1cm];
                    \draw[fill=light] (3,4) circle [radius=0.1cm];
                    \draw[fill=light] (4,3) circle [radius=0.1cm];
                \end{tikzpicture}
                \caption{Slices for $B^2_{2,2,1}$. Left side: $\mathcal{B}_{\mathfrak{m},\boldsymbol{\alpha}}$ of the topologically twisted theory for $\alpha(\mathfrak{m})=3$. Right side: $\mathcal{B}_{\mathfrak{m},\boldsymbol{\alpha}}$ of the exotic theory for $\alpha(\mathfrak{m})=3$. Only points in light blue contribute to the slices. The shape of the slice is as for $\mathbb{CP}^2$ \cite{Lundin:2021zeb} but, here, only certain pairs $(n_1,n_2)$ contribute.}
                \label{figure10}
            \end{figure}
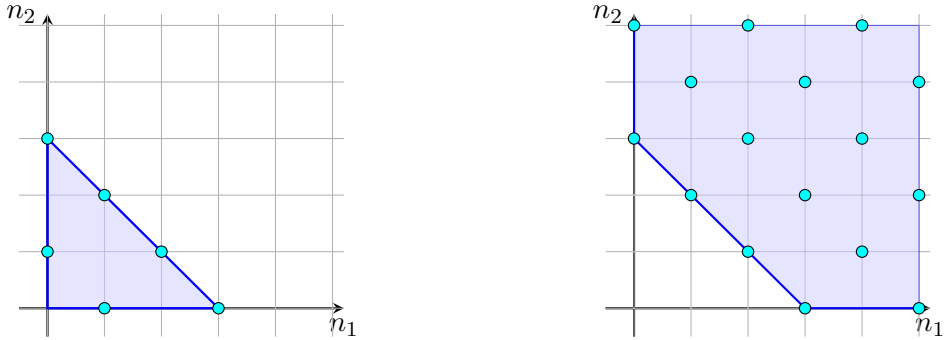

        \paragraph{Choice (ii).}\label{par}
            Let us for the moment set $k_1=1$ and return to the general case $k\neq1$ momentarily. Slices $\mathcal{B}_{t,\boldsymbol{\alpha}}\subset\mathbb{Z}^3_{\ge0}$ for $k_2=3,k_3=5$ and different values for $t$ are shown in \autoref{figure8}.                      
            For $h\to\infty$, only a single slice survives and contributes at the flux sector $t=\alpha(\mathfrak{m})$. We show the $(n_1,n_2)$-projection of these slices in \ref{figure9}.
            We can now understand the dependence on $k_1$. Let us recall that the slices $\mathcal{B}_{\mathfrak{m},\boldsymbol{\alpha}}$ are defined to be the pairs $(n_1,n_2)\in\mathbb{Z}^2_{\geq 0}$ satisfying:
            \begin{equation}\label{eq.k1neq1}
                k_3n_1+k_2n_2=t^{\text{top,ex}}\mp k_1n_3.
            \end{equation}
            If $k_1=1$ this condition translates into:
            \begin{equation}\begin{split}\label{eq.shape}
                \text{top}:\;&\quad k_3n_1+k_2n_2\leq t,\\
                \text{ex}:\;&\quad k_3n_1+k_2n_2\geq t.
            \end{split}\end{equation}
            Instead, if $k_1\neq 1$, only pairs $(n_1,n_2)$, for which there exists $n_3\in\mathbb{Z}_{\geq 0}$ such that \eqref{eq.k1neq1} holds, contribute. We present slices for $B^2_{2,2,1}$ in \autoref{figure10}.
            Unlike the previous case, one can check that $n_1$ (resp. $n_2$) jumps by $k_1$ at a fixed value of $n_2$ (resp. $n_1$) and thus the shape of the slice is not affected. More precisely, only pairs belonging to the sublattice of $\mathbb{Z}^2_{\geq 0}$ defined by
            \begin{equation}\label{eq.shifts2}
                \{(n_1,n_2)\in\mathbb{Z}^2_{\geq 0}\; |\; \exists c\in\mathbb{Z}_{\geq 0} : n_1+n_2=ck_1+\rmd{\alpha(\mathfrak{m})}{k_1}\}.
            \end{equation}
            and satisfying \eqref{eq.shape} contribute to the slice $\mathcal{B}_{\mathfrak{m},\boldsymbol{\alpha}}$. This explains, for example, why the origin does not contribute on the left-hand side in \autoref{figure10}.

            We have previously observed that both, $Z_{B^1_{\boldsymbol{\alpha}}}$ and $Z_{B^2_{\boldsymbol{\alpha}}}$ agree with the respective one-loop determinant on $\mathbb{CP}^1_{(N_1,N_2)}$ and $\mathbb{CP}^2_{(N_1,N_2,N_3)}$ for each flux sector if we choose the weights $\{N_i\}$ carefully. We propose that also for choice (ii) the one-loop determinant matches the one on an orbifold whose locus and degree of singularities agree with the branching structure of $S^5_{\boldsymbol\alpha}$. Rather than weighted projective space, for choice (ii) the corresponding orbifold is an orbifold projective space \cite{Abreu01}.

\section{Factorised Expressions}\label{sec--6}

    In the previous section we have computed the one-loop determinant for fixed $\mathfrak{m}$ and the classical part of the partition function on $B^2_{\boldsymbol{\alpha}}$. In this section we use equivariance under the respective torus-action to factorise the one-loop determinant into contributions from each torus fixed point (see e.g. \cite{Lundin:2023tzw} for more detail in the manifold case). The factorisation data can then be used to conjecture the instanton part of the partition function.

    It turns out that the local equivariance parameters obtained from factorisation for choice (i) match those on $\mathbb{CP}^2$ \cite{Lundin:2021zeb} with the only difference being an overall rescaling by $1/(\alpha_1\alpha_2\alpha_3)$ in front of the shift proportional to $\alpha(\mathfrak{m})$ and some shifts depending on $\rmd{\alpha(\mathfrak{m})}{\alpha_1}$ and $\rmd{\alpha(\mathfrak{m})}{\alpha_2}$. Thus, we focus on the factorisation of choice (ii) here. Let us start by expressing the five-dimensional perturbative partition function \eqref{eq.tperturbative.Salpha} as a product of contributions from neighbourhoods $\hat{\mathbb{C}}^2_{\epsilon_1,\epsilon_2}\times S^1$ around the fixed fibres (i.e. the vertices of the toric diagram; $\hat{\mathbb{C}}^2$ denotes the respective branched cover of $\mathbb{C}^2$). As for an ordinary $S^5$, the factorised expression for the perturbative partition function on $S^5_{\boldsymbol{\alpha}}$ is determined by a choice of imaginary part of the equivariance parameters\footnote{Each individual contribution is affected by the choice of imaginary part. However, the partition function is independent of this choice.} $\epsilon_1,\epsilon_2$, and thus of the vector $\X$. Then, we have the following factorisation property:
    \begin{equation}
        \prod_{\alpha\in\Delta}\prod_{t}\Upsilon^{\mathcal{B}_{ t,\boldsymbol{\alpha}}}\left(\ii\alpha(a) \pm\omega_3 t\bigg|\frac{\epsilon_1}{\alpha_1},\frac{\epsilon_2}{\alpha_2}\right)=\prod_{i=1}^3\prod_{\alpha\in\Delta} \prod_{t}\Upsilon_i( \ii\alpha(a) +\beta_i^{-1}t|\epsilon^i_1,\epsilon^i_2)^{s_i},
    \end{equation}
    where $i$ labels the three fixed fibres and the local equivariance parameters for the $T^3$-action are $\epsilon_1^i,\epsilon_2^i,\beta_i^{-1}$. These, together with the parameter $s_i$ can be read off from \autoref{tab--6}.
    \begin{table}[h]
        \centering
        \caption{Local equivariance parameters for the topological (left) and exotic (right) theories. For ease of notation we relabeled $\tilde{\beta}_i^{-1,\text{top}}\equiv\beta_i^{-1,\text{top}}-(k_1k_2k_3)^{-1}(1-1/3(\epsilon_1+\epsilon_2))$ in the topological and $\tilde{\beta}_i^{-1,\text{ex}}\equiv\beta_i^{-1,\text{ex}}-(3k_1k_2k_3)^{-1}(1-(\epsilon_1+\epsilon_2))$ in the exotic case.}\label{tab--6}
        \begin{tabular}{ c || c | c | c }
            $i$ & 1 & 2 & 3 \\ \hline
            & & & \\[\dimexpr-\normalbaselineskip+3pt]
            $\epsilon_1^{\text{top},i}$ & $\frac{\epsilon_1}{\alpha_1}$ & $-\frac{\epsilon_1}{\alpha_1}$ & $\frac{k_3\epsilon_1}{\alpha_1}-\frac{k_2\epsilon_2}{\alpha_2}$ \\ 
            $\epsilon_2^{\text{top},i}$ &  $\frac{\epsilon_2}{\alpha_2} $ & $ \frac{k_2\epsilon_2}{\alpha_2}-\frac{k_3\epsilon_1}{\alpha_1}$ & $-\frac{\epsilon_2}{\alpha_2} $ \\ 
            $\tilde{\beta}_i^{-1,\text{top}}$ & $0$ &  $\frac{\epsilon_1}{\alpha_1k_3}$ & $\frac{\epsilon_2}{\alpha_2k_2} $ \\
            $s_i$ & 1 & 1 & -1 \\
        \end{tabular}
        \hspace{0.5cm}
        \begin{tabular}{ c || c | c | c }
            $i$ & 1 & 2 & 3 \\ \hline
            & & & \\[\dimexpr-\normalbaselineskip+3pt]
            $\epsilon_1^{\text{ex},i}$ & $\frac{\epsilon_1}{\alpha_1}$ & $\frac{\epsilon_1}{\alpha_1}$ & $\frac{k_3\epsilon_1}{\alpha_1}-\frac{k_2\epsilon_2}{\alpha_2}$ \\ 
            $\epsilon_2^{\text{ex},i}$ &  $\frac{\epsilon_2}{\alpha_2} $ & $ \frac{k_2\epsilon_2}{\alpha_2}-\frac{k_3\epsilon_1}{\alpha_1}$ & $ \frac{\epsilon_2}{\alpha_2} $ \\ 
            $\tilde{\beta}_i^{-1,\text{ex}}$ & $0$ &  $\frac{\epsilon_1}{\alpha_1k_3}$ & $\frac{\epsilon_2}{\alpha_2k_2} $ \\
            $s_i$ & 1 & -1 & 1 \\
        \end{tabular}
    \end{table}
    The $\Upsilon_i$-functions are defined as follows:
    \begin{equation}
        \Upsilon_i(z|\epsilon^i_1,\epsilon^i_2)=\prod_{(j,k)\in\mathcal{D}_i}(\epsilon^i_1 j+\epsilon^i_2 k+z)\prod_{(j,k)\in\mathcal{D}_i^\prime}(\epsilon^i_1 j+\epsilon^i_2 k+\bar{z}).
    \end{equation}
    The lattices $\mathcal{D}_i,\mathcal{D}'_i\subset\mathbb{Z}^2$ depend on the imaginary part of $\epsilon_1,\epsilon_2$ and are closely related to those appearing in \cite{Lundin:2023tzw} for toric Sasakian manifolds. However, there are two main differences:
    \begin{itemize}
        \item When $t$ is not a multiple of either $k_3$ or $k_2$, the shifts, respectively, by $\beta_2^{-1},\beta_3^{-1}$ do not lead to points on the lattice. This happens because the distance from the shift by $\beta_1^{-1}$ is not integer.
        \item When $k_1\neq 1$ there is an extra complications as not all points in the regions $\mathcal{D}_i,\mathcal{D}'_i$ have to be considered, as it is obvious from \eqref{eq.shifts2}.
    \end{itemize}
    Thus, we define the regions $\mathcal{D}_i,\mathcal{D}'_i$ as in eq. (6.3) \cite{Lundin:2023tzw} but consider all values of\footnote{As in \cite{Lundin:2023tzw}, depending on the regularisation, $i,j$ can be greater than 0 or 1.} $(i,j)\in\mathbb{R}^2_{\geq 0}$ such that when they are shifted by $\beta_2^{-1},\beta_3^{-1}$ they belong to the sublattice defined in \eqref{eq.shifts2}.  

    Once the factorisation of the perturbative partition function on $S^5_{\boldsymbol{\alpha}}$ is determined, the one-loop contributions at each topological sector for $L^5_{\boldsymbol{\alpha}}(h,\pm 1)$ and $B^2_{\boldsymbol{\alpha}}$ follow simply by restricting $t$:
    \begin{equation}
        Z_{L^5_{\boldsymbol{\alpha}}(h,\pm 1)}=\prod_{\alpha\in\Delta}\prod_{i=1}^3 \prod_{t=\alpha(\mathfrak{m})\mmod h}\Upsilon_i( \ii\alpha(a) +\beta_i^{-1}t|\epsilon^i_1,\epsilon^i_2)^{s_i}
    \end{equation}
    \begin{equation}\label{eq--6.Bpert}
        Z_{B^2_{\boldsymbol{\alpha}}}=\prod_{\alpha\in\Delta}\prod_{i=1}^3 \Upsilon_i( \ii\alpha(a) +\beta_i^{-1}\alpha(\mathfrak{m})|\epsilon^i_1,\epsilon^i_2)^{s_i}.
    \end{equation}
     Note that, as for quasi-toric four-manifolds \cite{Lundin:2023tzw}, the flux-dependence enters via a shift of the Coulomb branch parameter $a$ which can be read out from \autoref{tab--6}.
     
     Finally, given that one-loop determinants agree between the branched covers $B^{1,2}_{\boldsymbol{\alpha}}$ and weighted projective space $\mathbb{CP}^{1,2}_{\boldsymbol{N}}$, the factorisation also agrees.

\section{Discussion}\label{sec--7}

    In this work we have computed the one-loop determinants at all flux sectors and classical action for the 2d $\mathcal{N}=(2,2)$ vector multiplet on $B^1_{\boldsymbol{\alpha}}$ (see \eqref{eq.1loop.spindle.top}-\eqref{eq.S2cl}) and the 4d $\mathcal{N}=2$ vector multiplet on $B^2_{\boldsymbol{\alpha}}$ (see \eqref{eq.CP2alphatop}-\eqref{eq.S4cl}). We obtained these results from dimensional reduction of the 3d $\mathcal{N}=2$ and 5d $\mathcal{N}=1$ vector multiplet on branched spheres $S^3$, resp. $S^5$. These are $S^1$-fibrations over $B^{1,2}_{\boldsymbol{\alpha}}$ and the reduction was implemented, according to \cite{Lundin:2023tzw}, by taking finite quotients along the $S^1$-fibre first. This introduced extra topological sectors which, in the limit where the order of the quotient group becomes infinite, label different flux sectors in the 2d (resp. 4d) theory. Depending on the choice of fibration we obtain a topologically twisted or exotic theory on the base space.

    \subsection{Weighted Projective Space \texorpdfstring{$\mathbb{CP}^{r-1}_{\boldsymbol{N}}$}{CPʳ⁻¹ₙ}}

        We have observed in previous sections the similarity of our results for the one-loop partition function on $B^1_{\boldsymbol{\alpha}}$ and $B^2_{\boldsymbol{\alpha}}$ with those on weighted projective spaces $\mathbb{CP}^1_{\boldsymbol{N}}$ and $\mathbb{CP}^2_{\boldsymbol{N}}$ for a suitable choice of $\boldsymbol{\alpha},\boldsymbol{N}$. This might seem perplexing, given that $B^{r-1}_{\boldsymbol{\alpha}}=\mathbb{CP}^{r-1}$ and, hence, one expects to get the standard theory on $\mathbb{CP}^{r-1}$. However, remember that our starting point for the dimensional reduction was the one-loop determinant on $S^{2r-1}_{\boldsymbol{\alpha}}$, and this result is obtained as some limit $\epsilon\to\infty$ of the one-loop determinant on a resolved space \cite{Nishioka:2013haa,Hama:2014iea} (see \autoref{app--geometry}). Taking the $S^1$-quotient of the latter along the locally free direction yields a base space which, in the limit $\epsilon\to\infty$, is precisely diffeomorphic to weighted projective space. Therefore, the agreement we observe is somewhat expected. 
        
        Leaving the argument of the resolved space aside, one might still wonder what theory on $\mathbb{CP}^{r-1}$ our one-loop partition function corresponds to. We observe that the main difference in partition functions, compared to the standard theory on $\mathbb{CP}^{r-1}$, is the contribution of fractional flux; the latter is concentrated at the boundary of the $(r-1)$-simplex (viewing $\mathbb{CP}^{r-1}$ as a $T^{r-1}$-fibration). Therefore, a natural guess is that our partition function corresponds to the theory on $\mathbb{CP}^{r-1}$ in the presence of defects supported on this boundary. This is corroborated by similar equivalences previously found between 2d (4d) gauge theories on smooth manifolds in the presence of Gukov-Witten defects and the corresponding gauge theory on an orbifold \cite{Kanno:2011fw,Hosomichi:2015pia}. We leave a precise identification of the defects necessary in our case to future work. 
        
        While we can only compare the one-loop partition function for the topological twist on $\mathbb{CP}^2_{\boldsymbol{N}}$ to previous results in the literature \cite{Martelli:2023oqk}, the discussion above motivates the following
        \begin{con}\label{conjecture}
            For a fixed flux sector $\mathfrak{m}$, the one-loop determinant for the 4d $\mathcal{N}=2$ vector multiplet for exotic theories agrees on the two spaces $B^2_{\boldsymbol{\alpha}}$ and $\mathbb{CP}^2_{\boldsymbol{N}}$,
            \begin{equation*}
                Z^\text{ex}_{B^2_{\boldsymbol{\alpha}}}(a,\mathfrak{m};\epsilon_1,\epsilon_2)=Z^\text{ex}_{\mathbb{CP}^2_{\boldsymbol{N}}}(a,\mathfrak{m};\epsilon_1,\epsilon_2),
            \end{equation*} 
            for $\gcd(\alpha_i,\alpha_j)=1$, $i=1,2,3$ and the weight vector $\boldsymbol{N}$ as in \eqref{eq.CP2.weights}.
        \end{con}
        \noindent 
        Note that we require $\boldsymbol{N}$ in terms of the branch indices $\boldsymbol{\alpha}$ to be such that the singularity degrees on $\mathbb{CP}^2_{\boldsymbol{N}}$ match precisely the branch indices on $S^5_{\boldsymbol{\alpha}}$ (see \autoref{fig--triangles}).

        Finally, we expect the conjecture to extend to topological and exotic theories on $B^2_{\boldsymbol{\alpha}}$ for \hyperref[par]{choice (ii)}, corresponding to orbifold projective spaces \cite{Abreu01}.

        \paragraph{Full Partition Function on $\mathbb{CP}^2_{\boldsymbol{N}}$.}
            Let us now comment on instantons in the theory on $\mathbb{CP}^2_{\boldsymbol{N}}$. By a standard argument \cite{Nekrasov:2003vi,Festuccia:2018rew}, the instanton part of the partition function is given by a product of Nekrasov partition functions on $\mathbb{C}^2_{\epsilon_1^i,\epsilon_2^i}/\mathbb{Z}_{N_i}$ \cite{Fucito:2004ry,Bonelli:2011jx,Bonelli:2011kv,Bonelli:2012ny,Bruzzo:2013nba,Bruzzo:2013daa} around the fixed points $i=1,2,3$. For each fixed point contribution, the Coulomb branch parameter is shifted by a corresponding flux contribution obtained from the factorisation. The instanton partition function is then obtained as
            \begin{equation}\label{eq--6.Binst}
                Z_{\mathbb{CP}^2_{\boldsymbol{N}}}^\text{inst}=\prod_{i=1}^\ell Z^{Nek}_{\mathbb{C}^2_{\epsilon_1^i,\epsilon_2^i}/\mathbb{Z}_{N_i}}(\ii a+\beta_i^{-1}\mathfrak{m}|\epsilon^i_1,\epsilon^i_2,q)\prod_{i=\ell+1}^3 Z^{Nek}_{\mathbb{C}^2_{\epsilon_1^i,\epsilon_2^i}/\mathbb{Z}_{N_i}}(\ii a+\beta_i^{-1}\mathfrak{m}|\epsilon^i_1,\epsilon^i_2,\bar{q}).
            \end{equation}
            The shifts in $a$ and the local equivariance parameters are discussed in \autoref{sec--6}. For the topological twist we have $\ell=3$ (i.e. instantons at all three fixed points) while for the exotic theory we have $\ell=2$ (i.e. anti-instantons at one fixed point). Consequently, the full partition function on weighted projective space $\mathbb{CP}^2_{\boldsymbol{N}}$ reads
            \begin{equation}\label{eq--7.full}
                \mathcal{Z}_{\mathbb{CP}^2_{\boldsymbol{\alpha}}}=\sum_{\mathfrak{m}}\int_{\mathfrak{h}}\dd a~e^{-S^\text{cl}_4/(\alpha_1\alpha_2\alpha_3)^2}\cdot Z_{B^2_{\boldsymbol{\alpha}}}\cdot Z_{\mathbb{CP}^2_{\boldsymbol{\alpha}}}^\text{inst},
            \end{equation}
            with $S^\text{cl}_4$ in \eqref{eq.S4cl} and $Z_{B^2_{\boldsymbol{\alpha}}}$ in \eqref{eq.CP2alphatop}, resp. \eqref{eq.CP2alphaex}. We believe a similar expression holds for orbifold projective spaces employing the shifts in $a$ and the equivariance parameters in \autoref{tab--6}.

    \subsection{Future Directions}
    
        \paragraph{Branched Coverings of Toric Sasakian Manifolds.} 
            A natural extension of this work is to consider the $\mathcal{N}=1$ vector multiplet on branched covers $M_{\boldsymbol{\alpha}}$ of other five-dimensional toric Sasakian manifolds $M$. We obtain these in a similar fashion by extending the periodicities of the three angles. We can again find locally free $S^1$-actions for these spaces along which we can reduce. The $\mathcal{B}_{t,\boldsymbol{\alpha}}$ involved in quotienting will now be slices not of $\mathbb{Z}_{\ge0}^{r}$ but of the dual moment map cone of $M$. With these modifications we could obtain the partition function on four-dimensional quasi-toric manifolds \cite{Festuccia:2019akm,Lundin:2023tzw}. Also in this case, we expect the one-loop determinant for a fixed flux sector to be equal to its orbifold pendant.
            
            The simplest examples to study are branched coverings of $Y^{p,q}$ \cite{Martelli:2004wu} which are $S^1$-fibrations over $B^1_{\boldsymbol{\alpha}}\times B^1_{\boldsymbol{\alpha}'}$. According to our findings, we expect that this matches the one-loop determinant including flux contributions on $\mathbb{CP}^1_{\boldsymbol{\alpha}}\times \mathbb{CP}^1_{\boldsymbol{\alpha}'}$.

        \paragraph{Locally Free $S^3$-Action.} 
            Weighted projective spaces are obtained quotienting $S^{2r-1}$ along a locally free $S^1$-action. Hence, orbifold singularities on the base space arise from a $\mathbb{Z}_{\alpha_i}$-subgroup of $U(1)$ acting trivially. Consequently, the singularities are of the form $\mathbb{C}^{r-1}/\mathbb{Z}_{\alpha_i}$. If we were instead to consider quotients by locally free $S^3$-actions, elements acting trivially can be given by finite subgroups of $SU(2)$. The simplest choice to investigate this is $S^7$. Here, the quotient along a locally free $S^3$-action yields weighted quaternionic projective space $\mathbb{HP}^1_{\boldsymbol{\alpha}}$, which is an $S^4$ with orbifold singularities at the two poles of the form $\mathbb{C}^{2}/\Gamma$ for a finite subgroup $\Gamma\subset SU(2)$. One could envision to obtain instantons in 4d from flat $SU(2)$-connections on finite quotients of $S^7$.

    \paragraph*{Acknowledgments}
        We are grateful to Guido Festuccia, Dario Martelli, Leonardo Santilli, Itamar Yaakov and Alberto Zaffaroni for stimulating discussions and correspondence on the subject. We thank Guido Festuccia for comments on the manuscript. RM acknowledges support from the Centre for Interdisciplinary Mathematics at Uppsala University. LR acknowledges support from the Shuimu Tsinghua Scholar Program.

\addtocontents{toc}{\protect\setcounter{tocdepth}{1}}
\appendix

\section{Branched Coverings and Weighted Projective Spaces}\label{app--geometry}


    In this appendix we briefly recall the definition of branched coverings and discuss the spaces $S^{3}_{\boldsymbol{\alpha}}$ and $S^5_{\boldsymbol{\alpha}}$ and their $S^1$-quotient. We also give the basic definition and properties of weighted projective space and point out similarities between the two spaces in support of \autoref{conjecture}.

    \subsection{Branched Coverings of Spheres}\label{subsec--2.branched.coverings}

        A map $p:\tilde{M}\rightarrow M$ into a $d$-dimensional manifold $M$ is called an $n$-fold \textit{branched covering} if there exists a codimension-2 subcomplex $L\subset M$, called the branch locus, such that $p^{-1}(L)\subset\tilde{M}$ is a codimension-2 subcomplex and $p|_{\tilde{M}-p^{-1}(L)}$ is an $n$-fold covering. A generic point $\tilde{y}\in p^{-1}(L)$ has an open neighbourhood homeomorphic to $D^2\times I^{d-2}$ on which $p$ takes the form $D^2\times I^{d-2}\rightarrow D^2\times I^{d-2}, (z,x)\mapsto (z^\alpha,x)$. The positive integer $\alpha$ is called the branch index of $\tilde{y}$ \footnote{The sum of the branch indices for all points in the preimage of a point in $y\in L$ is the degree of the cover, i.e. $\sum_{\tilde{y}\in p^{-1}(y)}\alpha(\tilde{y})=n$. This is generally true for any point $y\in M$, but only for $y\in L$ is the resulting partition of $n$ non-trivial.}. We can introduce polar coordinates $(r,\tilde\theta)$ with $\tilde\theta\sim\tilde\theta+2\pi q$ on the disc and write $\tilde z=r\e{\ii\tilde\theta}$ as local complex coordinate. Then $z=r^{1/q}\e{\ii\theta}$ with $\theta=\tilde\theta/q$ is $2\pi$-periodic; an illustration for $q=3$ is given in Figure \ref{fig.pest.branched}. Note that functions on the cover are smooth if they are smooth in $z$.
        \begin{figure}[htbp]
            \centering
            \begin{tikzpicture}[scale=.9]
                \draw[thick] (0,0) circle [radius=1.5cm];
                \fill (0,0) circle [radius=.07cm];
                \draw (0,0)--(1.5,0);
                \draw (0,0)--(-.75,1.5*.866);
                \draw (0,0)--(-.75,-1.5*.866);
                \fill (-1,0) circle [radius=.07cm] node [right]{$z_2$};
                \fill (.5,.866) circle [radius=.07cm] node [right]{$z_1$};
                \fill (.5,-.866) circle [radius=.07cm] node [left]{$z_3$};
                \draw[thick,-stealth] (2.5,0)--(4,0);
                \node at (3.25,0) [above]{$p$};
                \begin{scope}[shift={(6.5,0)}]
                    \draw[thick](0,0) circle [radius=1.5cm];
                    \fill (0,0) circle [radius=.07cm];
                    \fill (-1,0) circle [radius=.07cm] node [right]{$z$};
                \end{scope}
            \end{tikzpicture}
            \caption{Local map $z\mapsto z^3$ from the $\mathbb{C}$-plane transverse to the branch locus (the origin) down to the base.}
            \label{fig.pest.branched}
        \end{figure} 
        
        For our case, we take $M=S^{2r-1}$ and $L$ to be a codimension 2 subspace
        \begin{equation}
            L\subset\bigcup_{i=1}^r\{(z_1,\dots,z_r)\in S^{2r-1}|z_i=0\}.
        \end{equation}
        From the toric viewpoint which we take frequently, $S^{2r-1}$ can be seen as a Lagrangian $T^r$-fibration over the moment polytope which is an $(r-1)$-simplex. Then $L$ describes a subset of its facets (which are themselves $S^{2r-3}$). At intersections of two facets the corresponding spheres are glued along an $S^{2r-5}$ and so on. For $r>3$ this geometry becomes fairly complicated and since we focus on computing partition functions for $r=2,3$ we henceforth restrict to these low-dimensional cases. However, everything that follows can be extended beyond $r=3$. 

        \paragraph{Three Dimensions.}
            Consider a branched covering over the round $S^3$. We parametrise a point in $S^3$ (which we mostly view as a Hopf fibration over $S^2$ in the following) by
            \begin{equation}
                z_1=\sin\phi\e{\ii\theta_1},\qquad z_2=\cos\phi\e{\ii\theta_2}
            \end{equation}
            with angles $\phi\in[0,\pi/2]$ and $\Delta\theta_i=2\pi$. 
            In order to move to a branched covering $S^3_{\boldsymbol{\alpha}}$, we can simply extend the periodicity of the azimuthal angles to $\Delta\tilde{\theta}_i=2\pi\alpha_i$, which we now write as $\tilde{\theta}_i$, to distinguish from the canonical periodicity of $\theta_i$. 
            The branching can most easily be seen by viewing $S^3$ as a $T^2$-fibration over the 1-simplex; see Figure \ref{fig.pest.S3}. The torus for $S^3_{\boldsymbol{\alpha}}$ is an $\alpha_1\alpha_2$-fold regular cover over $T^2$. At $z_1=0$, the $a$-cycle degenerates and identifies $\alpha_1$ sheets, i.e., the $b$-cycle is part of the branch locus with index $\alpha_1$; it covers the standard Hopf fibre $\alpha_2$-fold. Similarly, at $z_2=0$, the $b$-cycle degenerates and the $a$-cycle becomes part of the branch locus with index $\alpha_2$, covering the Hopf fibre $\alpha_1$-fold. We have $\pi_1(S^3-L)\simeq \mathbb{Z}\times \mathbb{Z}$ (corresponding to the two cycles) and the monodromy $m$ maps the two generators to the ones of $\mathbb{Z}_{\alpha_1}\times \mathbb{Z}_{\alpha_2}$ accordingly.
            \begin{figure}[htbp]
                \centering
                \tikzset{
                    pics/torus/.style n args={3}{
                        code = {
                        \providecolor{pgffillcolor}{rgb}{1,1,1}
                        \begin{scope}[
                            yscale=cos(#3),
                            outer torus/.style = {draw,line width/.expanded={\the\dimexpr2\pgflinewidth+#2*2},line join=round},
                            inner torus/.style = {draw=pgffillcolor,line width={#2*2}}
                            ]
                            \draw[outer torus] circle(#1);\draw[inner torus] circle(#1);
                            \draw[outer torus] (180:#1) arc (180:360:#1);\draw[inner torus,line cap=round] (180:#1) arc (180:360:#1);
                        \end{scope}
                        }
                    }
                }
                \begin{tikzpicture}
                    \draw[thick] (0,0)--(5,0);
                    \fill (0,0) circle [radius=.07cm] node [below]{$\phi=0$};
                    \fill (5,0) circle [radius=.07cm] node [below]{$\phi=\frac{\pi}{2}$};
                    \pic[rotate=90,thick] at (2.5,1.5) {torus={.8cm}{2mm}{70}};
                    \draw[thick] (0,1.5) ellipse (.25cm and .05cm);
                    \draw[thick] (5,1.5) ellipse (.15cm and .7cm);
                \end{tikzpicture}
                \caption{$S^3$ represented as a torus fibration over the interval. At $\phi=0$, the $a$-cycle degenerates, while at $\phi=\pi/2$, the $b$-cycle degenerates.}
                \label{fig.pest.S3}
            \end{figure}
            
            For the round metric on $S^3$, the corresponding metric on $S^3_{\boldsymbol{\alpha}}$ is given (in terms of the $2\pi$-periodic angles) by
            \begin{equation}\label{eq--2.metric.3d}
                \dd s^2=\dd\phi^2+\alpha_1^2\sin^2\!\phi\,\dd\theta_1^2+\alpha_2^2\cos^2\!\phi\,\dd\theta_2^2.
            \end{equation}
            One can show that this space has conical singularities with surplus angle $2\pi\alpha_1$ at $\phi=0$ and $2\pi\alpha_2$ at $\phi=\pi/2$ \cite{Nishioka:2013haa}. 
            We are interested $S^1$-quotients of this space along the directions $\X^{\pm}=\partial_{\tilde{\theta}_1}\pm\partial_{\tilde{\theta}_2}$, with respect to which we can view $S^3_{\boldsymbol{\alpha}}$ as a ``branched (anti-)Hopf fibration''.


        \paragraph{Five Dimensions.}
            Let us start with parametrising the round $S^5$ by
            \begin{equation}
                z_1=\sin\varphi\cos\phi\e{\ii\theta_1},\qquad z_2=\sin\varphi\sin\phi\e{\ii\theta_2},\qquad z_3=\cos\varphi\e{\ii\theta_3},
            \end{equation}
            with angles $\varphi,\phi\in[0,\pi/2]$ and $\Delta\theta_i=2\pi$ for $i=1,2,3$. Similarly to the three-dimensional case, in order to move to a branched cover $S^5_{\boldsymbol{\alpha}}$ we simply extend the periodicity of the azimuthal angles to $\Delta\tilde{\theta}_i=2\pi\alpha_i$. 
            Again, we can view $S^5$ as a $T^3$-fibration over the 2-simplex. Then the torus for $S^5_{\boldsymbol{\alpha}}$ is an $\alpha_1\alpha_2\alpha_3$-fold cover over $T^3$. At $z_1=0$ (corresponding to an edge of the simplex), the $a$-cycle degenerates and identifies $\alpha_1$ sheets, i.e., the corresponding three-dimensional subspace is an $\alpha_2\alpha_3$-fold cover of $S^3$ which is part of the branch locus and has index $\alpha_1$. Similarly, the other two edges of the simplex are associated to an $\alpha_1\alpha_3$-fold cover of $S^3$ with branch index $\alpha_2$ and an $\alpha_1\alpha_2$-fold cover of $S^3$ with index $\alpha_3$. 
            At the vertex where $a$ and $b$-cycles of the torus degenerate, the corresponding subspace in $S^5_{\boldsymbol{\alpha}}$ is an $\alpha_3$-fold cover of $S^1$ with branch index $\alpha_1\alpha_2$ and similar for the other two vertices.
            Hence, the branch locus $L$ in this case is stratified and consists of three $S^3$, which intersect pairwise in an $S^1$ corresponding to a vertex of the 2-simplex (see \autoref{fig--triangles}).

            For the round metric on $S^5$, the corresponding metric on $S^5_{\boldsymbol{\alpha}}$ is given (in terms of the $2\pi$-periodic angles) by
            \begin{equation}
                \dd s^2=\dd\varphi^2+\sin^2\!\varphi\,\dd\phi^2+\alpha_3^2\cos^2\!\varphi\dd\theta_3^2+\alpha_2^2\sin^2\!\varphi\sin^2\!\phi\,\dd\theta_2^2+\alpha_1^2\sin^2\!\varphi\cos^2\!\phi\,\dd\theta_1^2.
            \end{equation}
            Similarly to the three-dimensional case, one can show that this space has conical singularities at the branch locus with surplus angles corresponding to the respective branch index.
            We are interested in $S^1$-quotients of this space along the directions $\X^{\pm}=\partial_{\tilde{\theta}_1}+\partial_{\tilde\theta_2}\pm\partial_{\tilde{\theta}_3}$, with respect to which we can view $S^5_{\boldsymbol{\alpha}}$ as a ``branched (anti-)Hopf fibration''.

    \subsection{Weighted Projective Space from Spheres}\label{subsec--2.locally.free.actions}
    
        Hereinafter, we consider odd-dimensional spheres $S^{2r-1}$ embedded into $\mathbb{C}^{r}$ in the standard way. In particular, we consider $S^{2r-1}$ as an $S^1$-fibration for the following \textit{locally free} $S^1$-action:
        \begin{equation}\label{eq--2.locally.free.action}
            (z_1,\dots,z_{r})\mapsto(\e{\ii N_1\xi}z_1,\dots,\e{\ii N_r\xi}z_r),\qquad \xi\sim \xi+2\pi,
        \end{equation}
        where $N_1,\dots,N_r\in\mathbb{N}$ with $\gcd(N_1,\dots,N_r)=1$ (this guarantees an effective action). We commonly refer to these as weights and define the weight vector $\boldsymbol{N}=(N_1,\dots,N_r)$. The locally free nature of this action can be seen easily by restricting to the $S^1$-fibres $|z_i|=1$ ($i=1,\dots,r$). There, group elements of the form $\xi=2\pi k/N_i$ for $k\in\mathbb{Z}_{N_i}$ act trivially. Consequently, the base space of the fibration contains orbifold singularities and, more precisely, we obtain an $S^1$-orbibundle
        \begin{equation}\label{eq--2.fibration}
            \begin{tikzcd}[column sep=small]
                S^1\ar[r] & S^{2r-1}\ar[d,"\pi"]\\
                & \mathbb{CP}^{r-1}_{\boldsymbol{N}}
            \end{tikzcd}
        \end{equation}
        over a weighted projective space\footnote{This definition of complex weighted projective space might be unfamiliar to some readers who instead have in mind the definition as the symplectic quotient $\mathbb{C}^r/\!/S^1$ by \eqref{eq--2.locally.free.action}. Topologically, the two procedures agree. In this article, we always understand $\mathbb{CP}^{r-1}_{\boldsymbol{N}}$ in the sense of \eqref{eq--2.fibration}.} $\mathbb{CP}^{r-1}_{\boldsymbol{N}}$. Note that $\mathbb{CP}^{r-1}_{(1,\dots,1)}=\mathbb{CP}^{r-1}$ and \eqref{eq--2.fibration} is simply the Hopf fibration. This case has been studied in \cite{Lundin:2021zeb}.
    
        The orbifold structure of $\mathbb{CP}^{r-1}_{\boldsymbol{N}}$ can be inferred from \eqref{eq--2.locally.free.action} as follows: denote by $[z]$ a point in $\mathbb{CP}^{r-1}_{\boldsymbol{N}}$ with $z_j\neq0$ for some $j\in\{1,\dots,r\}$ and let $m$ be the greatest common divisor of the corresponding set of weights $N_j$. Then the isotropy group $\Gamma_{[z]}$ of $[z]$ is isomorphic to $\mathbb{Z}_m$. In particular, we have $\Gamma_{[z]}\simeq\mathbb{Z}_{N_i}$ for $[z]=[0:\hdots:0:z_i:0:\hdots:0]$ with $z_i\neq0$ and, due to $\gcd\boldsymbol{N}=1$, at points where $z_i\neq0$ for all $i\in\{1,\dots,r\}$, $\Gamma_{[z]}$ is trivial.

        \paragraph{Three Dimensions.} 
            In this case, \eqref{eq--2.fibration} simply corresponds to a Seifert fibration of $S^3$ with two exceptional fibres $(N_1,\beta_1)$, $(N_2,\beta_2)$ located at $|z_1|=1$ and $|z_2|=1$, respectively. In order for the fundamental group of the Seifert fibration to be trivial, we impose\footnote{This can always be arranged for given $N_1,N_2$ with $\gcd(N_1,N_2)=1$ by virtue of Bezout's lemma.} $N_1\beta_2+N_2\beta_1=1$. The base space, which topologically is $S^2$, has conical singularities at the two poles with deficit angle $2\pi/N_1$ and $2\pi/N_2$, respectively. This space is commonly referred to as a spindle. 
    
            The orbifold fundamental group of $\mathbb{CP}^1_{\boldsymbol{N}}$ is trivial, its second integral cohomology is $H^2(\mathbb{CP}^1_{\boldsymbol{N}})\simeq\mathbb{Z}$ \cite{Kawasaki:1973} and its (orbifold) Euler characteristic is $\chi(\mathbb{CP}^1_{\boldsymbol{N}})=\frac{1}{N_1}+\frac{1}{N_2}$.
    
        \paragraph{Five Dimensions.} 
            In this case, the singularity structure of \eqref{eq--2.fibration} can be more interesting. Viewing $\mathbb{CP}^2_{(N_1,N_2,N_3)}$ as a toric orbifold for the moment, the $i$th facet $F_i$ of its polytope is located at $z_i=0$. Hence, points in the interior $\mathring{F}_i$ have isotropy group isomorphic to $\mathbb{Z}_{\gcd(N_j,N_k)}$, where $j,k\neq i$. Moreover, the facet $F_i$ itself corresponds to a spindle $\mathbb{CP}^1_{(N_j,N_k)}$. At a vertex of the polytope where facets $F_i,F_j$ intersect, we have isotropy group isomorphic to $\mathbb{Z}_{N_k}$, where $i,j\neq k$. The singularity structure is displayed in \autoref{fig--triangles}.
            \begin{figure}
                \centering
                \begin{tikzpicture}[scale=.85]
                    \draw[line width=1.2pt] (0,0) node[below left]{$N_3$}--(4,0) node[below right]{$N_1$}--(0,4) node[above]{$N_2$}--cycle;
                    \node at (2,0) [below,align=center]{$\gcd(N_1,N_3)$};
                    \node at (0,2) [left,align=center]{$\gcd(N_2,N_3)$};
                    \node at (2.5,2) [right,align=center]{$\gcd(N_1,N_2)$};
                    \begin{scope}[shift={(9,0)}]
                        \draw[line width=1.2pt] (0,0) node[below left]{$\alpha_1\alpha_2;\alpha_3$}--(4,0) node[below right]{$\alpha_2\alpha_3;\alpha_1$}--(0,4) node[above]{$\alpha_1\alpha_3;\alpha_2$}--cycle;
                        \node at (2,0) [below,align=center]{$\alpha_2;\alpha_1\alpha_3$};
                        \node at (0,2) [left,align=center]{$\alpha_1;\alpha_2\alpha_3$};
                        \node at (2.5,2) [right,align=center]{$\alpha_3;\alpha_1\alpha_2$};
                    \end{scope}
                \end{tikzpicture}
                \caption{Left-hand side: singularity structure of $\mathbb{CP}^2_{\boldsymbol{N}}$ in terms of the $S^5$ moment polytope. Right-hand side: branching structure of $S^5_{\boldsymbol{\alpha}}$ in terms of the $S^5$-moment polytope. The $(\alpha_i;\alpha_j\alpha_k)$-labelled edge corresponds to an $\alpha_j\alpha_k$-fold cover of $S^3$ of branch index $\alpha_i$, while the $(\alpha_i\alpha_j;\alpha_k)$-labelled vertex corresponds to an $\alpha_k$-fold cover of $S^1$ of branch index $\alpha_i\alpha_j$.}
                \label{fig--triangles}
            \end{figure}
            As an example, let us consider the weight vector $\boldsymbol{N}=(k_1k_2,k_1k_3,k_2k_3)$, for some positive integers $k_1,k_2,k_3$ with $\gcd(k_i,k_j)=1$ for $i,j=1,2,3$.
            Then we have 
            \begin{equation}
                \Gamma_{[z]\in\mathring{F}_1}\simeq\mathbb{Z}_{k_3},\qquad\Gamma_{[z]\in\mathring{F}_2}\simeq\mathbb{Z}_{k_2},\qquad\Gamma_{[z]\in\mathring{F}_3}\simeq\mathbb{Z}_{k_1}.
            \end{equation}
            For the facets themselves we have $F_1\simeq\mathbb{CP}^1_{(k_1,k_2)}$, $F_2\simeq\mathbb{CP}^1_{(k_3,k_1)}$, $F_3\simeq\mathbb{CP}^1_{(k_3,k_2)}$ and hence, at the vertices:
            \begin{equation}\label{eq--2.CP2sing}
                \Gamma_{F_1\cap F_2}\simeq\mathbb{Z}_{k_2k_3},\qquad\Gamma_{F_2\cap F_3}\simeq\mathbb{Z}_{k_1k_2},\qquad\Gamma_{F_1\cap F_3}\simeq\mathbb{Z}_{k_1k_3}.
            \end{equation}
            Hence, for $r\ge3$ we have, in general, already codimension 2 singularities.
    

    \subsection{Quotients of Branched Spheres}\label{subsec--2.quotients}

        In the main text we compute partition functions corresponding to a 2d, respectively 4d pure gauge theory by taking the limit $h\to\infty$. Here, we briefly discuss the quotient space $B^{r-1}_{\boldsymbol{\alpha}}$ ($r=1,2$) on which these theories are defined. Note that the quotient with respect to the locally free $S^1$-actions generated by $\X^\pm$ only differ by their relative orientation. We therefore only discuss the quotient corresponding to $\X^+$.

        \paragraph{Three Dimensions.} 
            Given the branched three-sphere $S^3_{\boldsymbol{\alpha}}$ with metric \eqref{eq--2.metric.3d}, we want to compute the metric fibration with respect to the locally free direction $\X^+=\partial_{\tilde\theta_1}+\partial_{\tilde\theta_2}$ (remember $\Delta\tilde\theta_i=2\pi\alpha_i$). For this purpose, we define new angles
            \begin{equation}
                \psi=\alpha_2n_1\tilde\theta_1+\alpha_1n_2\tilde{\theta}_2,\qquad\xi=\tilde\theta_2-\tilde\theta_1
            \end{equation}
            with integers $n_1,n_2$ such that $n_1\alpha_2+n_2\alpha_1=1$. The periodicities of the new angles are $\Delta\psi=2\pi\alpha_1\alpha_2$ and $\Delta\xi=2\pi$ and the metric of $S^3_{\boldsymbol{\alpha}}$ can be expressed as
            \begin{equation}
                \dd s^2=\frac{1}{4}\left(\dd\tilde\phi^2+\sin^2\!\tilde\phi\,\dd\xi^2\right)+\left(\dd\psi+(\alpha_2n_1\sin^2\tfrac{\tilde\phi}{2}-\alpha_1n_2\cos^2\tfrac{\tilde\phi}{2})\dd\xi\right)^2,
            \end{equation}
            where $\tilde\phi=2\phi$. Thus, we see that the base space is simply $S^2$.
            
            The theory on $S^3_{\boldsymbol{\alpha}}$ is commonly described as the limit where $\epsilon\to\infty$ of a theory on the resolved space \cite{Nishioka:2013haa}
            \begin{equation}
                \dd s^2_\text{res}=f_\epsilon(\phi)\dd\phi^2+\sin^2\!\phi\,\dd\tilde\theta_1^2+\cos^2\!\phi\,\dd\tilde\theta_2^2,
            \end{equation}
            where $f_\epsilon$ is a smooth function such that
            \begin{equation}
                f_\epsilon(\phi)=\begin{cases}
                    \alpha_1^2, & \phi=0\\\alpha_2^2, & \phi=\tfrac{\pi}{2}\\ 1, & \epsilon<\phi<\tfrac{\pi}{2}-\epsilon
                \end{cases}.
            \end{equation}
            Quotienting this space along $\X^+$ yields the following metric on the base:
            \begin{equation}
                \dd s^2_{res,B}=\frac{1}{4}\left(f_\epsilon(\tilde\phi)\dd\tilde\phi^2+\sin^2\!\tilde\phi\,\dd\xi^2\right).
            \end{equation}
            Remember that $\xi$ has periodicity $2\pi$, hence, the base space has conical singularities of degree $\alpha_1$ at $\tilde\phi=0$ and $\alpha_2$ at $\tilde\phi=\pi$. In other words, this space is diffeomorphic to a spindle $\mathbb{CP}^1_{(\alpha_1,\alpha_2)}$.

            Finally, let us comment on finite quotients by $\mathbb{Z}_h$ along $\X^+$. In the main text, we assume $\gcd(h,\alpha_i)=1$. The resulting space $L_{\boldsymbol{\alpha}}(h,1)\equiv S^3_{\boldsymbol{\alpha}}/\mathbb{Z}_h$ is again a branched cover with a branch locus $S^1\sqcup S^1$, but this time over the lens space $L(h,1)$. If, instead, we were to choose $h=k\alpha_1\alpha_2$ with $k\in\mathbb{N}$ and $\gcd(k,\alpha_i)=1$, then $S^3_{\boldsymbol{\alpha}}/\mathbb{Z}_h=L(k,1)$. Finally, choosing $h=k\alpha_1$ (with $k$ as above) undoes the branching on one $S^1$ but keeps it on the other $S^1$, so that the resulting space is a branched lens space $L_{(1,\alpha_2)}(k,1)$. In the limit where $h\to\infty$, all these choices yield $S^2$ as the base space.

        \paragraph{Five Dimensions.} 
            In five dimensions, the fibration along $\X^+=\partial_{\tilde\theta_1}+\partial_{\tilde\theta_2}+\partial_{\tilde\theta_3}$ works in complete analogy to the 3d case. Here, we consider $\gcd(\alpha_i,\alpha_j)=1$ for all $i\neq j$ (called choice (i) in the main text), for which the base space is simply $\mathbb{CP}^2$. A resolved space for $S^5_{\boldsymbol{\alpha}}$ is given by
            \begin{equation}
                \dd s^2_\text{res}=f_\epsilon(\varphi)\dd\varphi^2+g_\varepsilon(\phi)\sin^2\!\varphi\,\dd\phi^2+\cos^2\!\varphi\dd\tilde\theta_3^2+\sin^2\!\varphi\sin^2\!\phi\,\dd\tilde\theta_2^2+\sin^2\!\varphi\cos^2\!\phi\,\dd\tilde\theta_1^2,
            \end{equation}
            where $f_\epsilon,g_\varepsilon$ are smooth functions such that
            \begin{equation}
                f_\epsilon(\varphi)=\begin{cases}
                    \alpha_3^2, & \varphi=\pi/2\\
                    1, & 0\le\varphi<\pi/2-\epsilon
                \end{cases}~,\qquad
                g_\varepsilon(\phi)=\begin{cases}
                    \alpha_2^2, & \phi=0\\
                    \alpha_1^2, & \phi=\pi/2\\
                    1, & \tilde{\epsilon}<\phi<\pi/2-\tilde\epsilon
                \end{cases}.
            \end{equation}
            Quotienting this space along $\X^+$ yields the following metric on the base:
            \begin{equation}
                \dd s^2_{\text{res},B}=f_\epsilon(\varphi)\dd\varphi^2+\frac{1}{4}\sin^2\varphi(g_\varepsilon(\tilde\phi)\dd\tilde\phi^2+\sin^2\tilde\phi^2\dd\gamma^2)+\sin^2\varphi\cos^2\varphi(\dd\beta+\cos\tilde\phi\dd\gamma)^2,
            \end{equation}            
            where $\beta\equiv\tilde\theta_3-\tilde\theta_1-\tilde\theta_2$ and $\gamma\equiv\tilde\theta_2-\tilde\theta_1$ are both $2\pi$-periodic. Thus, the base space has codimension-two conical singularities at $\varphi=\frac{\pi}{2}$ of degree $\alpha_3$, at $\tilde\phi=0$ of degree $\alpha_2$ and at $\tilde\phi=\pi$ of degree $\alpha_1$. Moreover, it has codimension-four singularities at $\varphi=0$ of degree $\alpha_1\alpha_2$, at $\varphi=\frac{\pi}{2},\tilde\phi=0$ of degree $\alpha_2\alpha_3$ and at $\varphi=\frac{\pi}{2},\tilde\phi=\pi$ of degree $\alpha_1\alpha_3$. In other words, this space is diffeomorphic to weighted projective space $\mathbb{CP}^2_{\boldsymbol{N}}$, where $N=(\alpha_2\alpha_3,\alpha_1\alpha_3,\alpha_1\alpha_2)$.

\section{Dimensional Reduction of the Spindle Index}\label{app--A}

    In this appendix we show how to match our result for the partition function of the 2d vector multiplet on the branched cover $B^1_{\boldsymbol{\alpha}}$ with that on a spindle $\mathbb{CP}^1_{\boldsymbol{\alpha}}$. The latter is obtained from dimensional reduction of the spindle index computed in \cite{Inglese:2023wky}, to which we refer for details. Note that our results on the spindle are obtained reducing from a single theory for an $\mathcal{N}=2$ vector multiplet on the branched cover of the sphere $S^3_{\boldsymbol{\alpha}}$. Instead, the dimensional reduction of the spindle index, as it is performed along a trivially fibred $S^1$, it is such that from each $\mathcal{N}=2$ theory on $\mathbb{CP}^1_{\boldsymbol{\alpha}}\times S^1$ one obtains a unique theory on $\mathbb{CP}^1_{\boldsymbol{\alpha}}$. 
    
    \subsection{Spindle Index}
        
        Following \cite{Inglese:2023wky}, the partition function on $\mathbb{CP}^1_{\boldsymbol{\alpha}}\times S^1$ for an $\mathcal{N}=2$ vector multiplet can be written as:
        \begin{equation}\label{eq.spindleindex}
            \mathcal{Z}_{\mathbb{CP}^1_{\boldsymbol{\alpha}}\times S^1}^\sigma=\sum_{\mathfrak{m}}\oint \dd u\, e^{-S_{cl}}\cdot Z^{vm,\sigma}_{\mathbb{CP}^1_{\boldsymbol{\alpha}}\times S^1}.
        \end{equation}
        The sum is over magnetic charges $\mathfrak{m}$ that label configurations with flux for the field strength on the spindle. The classical part is given by a Chern-Simons term evaluated on the BPS locus,
        \begin{equation}
            S_\text{cl}=\frac{2\pi\ii k}{\alpha_1\alpha_2}\tr(\mathfrak{m}u),
        \end{equation}
        and the integral is performed over the holonomies $u$ for the gauge connections. Topologically twisted and exotic theories correspond, respectively, to $\sigma=1$ and $\sigma=-1$. For convenience, let us start by introducing some notation from \cite{Inglese:2023wky}:
        \begin{equation}\begin{split}\label{eq.definitions}
            &\mathfrak{m}=\alpha_1m_--\alpha_2m_+,\qquad p_+=\alpha(m_+)-\sigma,\qquad p_-=\alpha(m_-)+1,\\
            &\mathfrak{b}=1+\sigma\lfloor\sigma p_+/\alpha_1\rfloor+\lfloor-p_-/\alpha_2\rfloor,\qquad\mathfrak{c}=\rmd{-p_-}{\alpha_2}/\alpha_2-\sigma\,\rmd{p_+}{\alpha_1}/\alpha_1,\\
            &\gamma=-\frac{n}{2}+\omega\frac{\chi_{-\sigma}}{4},\qquad \frac{\chi_\sigma}{2}=\frac{1}{2}\left(\frac{1}{\alpha_2}+\frac{\sigma}{\alpha_1}\right)=\frac{1}{2\pi}\int_{\mathbb{CP}^1_{\boldsymbol{\alpha}}}\dd A_R,\\
            &q=e^{2\pi\ii\omega},\qquad y=q^{\mathfrak{c}/2}e^{2\pi\ii(2\gamma-\alpha(u))},
        \end{split}\end{equation}
        where $m_{\pm}\in\mathbb{Z}$, $n\in\mathbb{Z}$ and $\alpha$ denote the roots of the gauge algebra. The notation $\lfloor x\rfloor$ denotes the floor function and $\sigma=\pm1$ for the topologically twisted and exotic theory, respectively (or the twist and anti-twist in the language of \cite{Inglese:2023wky}).
        
        The one-loop contribution to the partition function for the 3d vector multiplet on $\mathbb{CP}^1_{\boldsymbol{\alpha}}\times S^1$ can be obtained from an index computation of a twisted Dolbeault operator and the index reads
        \begin{equation}\begin{split}
            I^\sigma_{\mathbb{CP}^1_{\boldsymbol{\alpha}}}=&\frac{1}{\alpha_1}\sum_{j=0}^{\alpha_1-1}\frac{\omega_+^{-jp_+}q_+^{-p_+}}{1-\omega_+^{j\sigma}q_+^\sigma}+\frac{1}{\alpha_2}\sum_{j=0}^{\alpha_2-1}\frac{\omega_-^{-jp_-}q_-^{-p_-}}{1-\omega_-^{j\sigma}q_-^\sigma}\\
            =&\frac{q^{-\sigma\lfloor\sigma p_+/\alpha_1\rfloor}}{1-q^\sigma}+\frac{q^{\lfloor -p_-/\alpha_2\rfloor}}{1-q^{-1}}.
        \end{split}\end{equation}
        Here, $q_\pm=q^{1/\alpha_{1,2}}$ and $\omega_\pm=e^{2\pi\ii/\alpha_{1,2}}$. This result can be combined with the one on the $S^1$ and the fugacities for gauge and flavour symmetries and then converted into the one-loop determinant:
        \begin{equation}\label{eq.spindleS1.oneloop}
            Z^{\sigma}_{\mathbb{CP}^1_{\boldsymbol{\alpha}}\times S^1}=\prod_{\alpha}(-y)^{\frac{1}{4}(1-\sigma-2\mathfrak{b})}q^{\frac{1}{8}(1-\sigma)(\mathfrak{b}-1)}\frac{\left(q^{\frac{1}{2}(1+\mathfrak{b})}y^{-1};q\right)_\infty}{\left(q^{\frac{\sigma}{2}(1-\mathfrak{b})}y^{-\sigma};q\right)_\infty},
        \end{equation}
        where $(z;q)_n$ is the $n$-th $q$-Pochhammer symbol
        \begin{equation}
            (a;q)_n=\prod_{k=0}^{n-1}\left(1-aq^k\right).
        \end{equation}

    \subsection{Dimensional Reduction}
        
        In order to match with our results in 2d we compute the small radius-limit where the $S^1$-factor shrinks. Our computation follows the dimensional reduction for the superconformal index in \cite{Benini:2012ui}. We set
        \begin{equation}
            q=e^{-\beta},\quad u=\frac{\beta a}{2\pi},
        \end{equation}
        where $\beta$ is the circumference of the $S^1$ and $a$ the 2d scalar arising from the component of the gauge connection along the $S^1$. This gives:
        \begin{equation}
           y=e^{-\beta\left(\frac{\mathfrak{c}}{2}+r\frac{\chi_{-\sigma}}{4}+\ii\alpha(a)\right)},
        \end{equation}
        as $r\in 2\mathbb{Z}$ and $n\in\mathbb{Z}$. In the limit $\beta\rightarrow 0$ the following relation holds:
        \begin{equation}
            \lim_{z\rightarrow 1}\frac{(z^s;z)_{\infty}}{(z^t;z)_\infty}=\frac{\Gamma(t)}{\Gamma(s)}(1-z)^{t-s},
        \end{equation}
        where
        \begin{equation}
            \Gamma(q)=\frac{\sqrt{2\pi}}{\prod_{n=0}^\infty(q+n)}.
        \end{equation}
        In our case $z=e^{-\beta}$ and
        \begin{equation}\begin{split}
            s=&\frac{1}{2}\left(1+\mathfrak{b}-\mathfrak{c}-r\chi_{-\sigma}/2-\ii\alpha( a)\right),\\
            t=&\frac{\sigma}{2}\left(1-\mathfrak{b}-\mathfrak{c}-r\chi_{-\sigma}/2-\ii\alpha( a)\right).
        \end{split}\end{equation}
        Hence, simply using $\lim_{\beta\to 0}(e^{-\beta})=1+\mathcal{O}(\beta)$, all contributions to the left of the Pochhammer symbols in \eqref{eq.spindleS1.oneloop} simplify and we find:
        \begin{equation}\label{eq.1loop.spindle}
            Z^{\sigma}_{\mathbb{CP}^1_{\boldsymbol{\alpha}}}=\lim_{\beta\rightarrow 0}Z^{\sigma}_{\mathbb{CP}^1_{\boldsymbol{\alpha}}\times S^1}\beta^{\sum_{\alpha}(s-t)}=\prod_{\alpha}\prod_{k=0}^\infty\frac{(s+k)}{(t+k)}.
        \end{equation}
        This result is the one-loop at each flux sector on $\mathbb{CP}^1_{\boldsymbol{\alpha}}$. The extra power of $\beta$, which is needed to recover the result on $S^2$ starting from the superconformal index with $\sigma=-1$ for $\alpha_1=\alpha_2=1$, is discussed in section 8.1 of \cite{Benini:2012ui}.

    \subsection{Spindle}
    
        First we rewrite
        \begin{equation}\begin{split}
            \mathfrak{b}=&1+\sigma\lfloor\sigma p_+/\alpha_1\rfloor+\lfloor-p_-/\alpha_2\rfloor\\
            =&1+\frac{\alpha(m_+)}{\alpha_1}-\frac{\alpha(m_-)}{\alpha_2}-\sigma\,\frac{\rmd{\sigma p_+}{\alpha_1}}{\alpha_1}-\frac{\rmd{-p_-}{\alpha_2}}{\alpha_2}-\left(\frac{1}{\alpha_2}+\frac{\sigma}{\alpha_1}\right)\\
            =&1-\frac{\alpha(\mathfrak{m})}{\alpha_1\alpha_2}-\sigma\,\frac{\rmd{\sigma p_+}{\alpha_1}}{\alpha_1}-\frac{\rmd{-p_-}{\alpha_2}}{\alpha_2}-\left(\frac{1}{\alpha_2}+\frac{\sigma}{\alpha_1}\right).
        \end{split}\end{equation}
        Also recalling
        \begin{equation}
            \mathfrak{c}=\frac{\rmd{-p_-}{\alpha_2}}{\alpha_2}-\sigma\,\frac{\rmd{p_+}{\alpha_1}}{\alpha_1},
        \end{equation}
        we find that
        \begin{equation}\begin{split}\label{eq.s.generic}
            \frac{1}{2}\left(1+\mathfrak{b}-\mathfrak{c}-\chi_{-\sigma}\right)&=\frac{1}{2}\left(2-\frac{\alpha(\mathfrak{m})}{\alpha_1\alpha_2}-2\,\frac{\rmd{-p_-}{\alpha_2}}{\alpha_2}-\left(\frac{1}{\alpha_2}+\frac{\sigma}{\alpha_1}\right)-\left(\frac{1}{\alpha_2}-\frac{\sigma}{\alpha_1}\right)\right)\\
            &=1-\frac{\alpha(\mathfrak{m})}{2\alpha_1\alpha_2}-\frac{\rmd{-p_-}{\alpha_2}}{\alpha_2}-\frac{1}{\alpha_2},
        \end{split}\end{equation}
        and
        \begin{equation}\begin{split}\label{eq.t.generic}
            \frac{\sigma}{2}\left(1-\mathfrak{b}-\mathfrak{c}-\chi_{-\sigma}\right)&=\frac{\sigma}{2}\left(\frac{\alpha(\mathfrak{m})}{\alpha_1\alpha_2}+2\sigma\,\frac{\rmd{\sigma p_+}{\alpha_1}}{\alpha_1}+\left(\frac{1}{\alpha_2}+\frac{\sigma}{\alpha_1}\right)-\left(\frac{1}{\alpha_2}-\frac{\sigma}{\alpha_1}\right)\right)\\
            &=\sigma\left(\frac{\alpha(\mathfrak{m})}{2\alpha_1\alpha_2}+\sigma\,\frac{\rmd{\sigma p_+}{\alpha_1}}{\alpha_1}\right)+\frac{1}{\alpha_1}.
        \end{split}\end{equation}
        The two expressions above simplify noting that
        \begin{equation}\begin{split}\label{eq.st.vector}
            &-\frac{\rmd{-p_-}{\alpha_2}}{\alpha_2}=-1+\frac{1}{\alpha_2}+\frac{\rmd{\alpha(m_-)}{\alpha_2}}{\alpha_2},\\
            &\sigma\,\frac{\rmd{\sigma p_+}{\alpha_1}}{\alpha_1}=\sigma\left(1-\frac{1}{\alpha_1}-\frac{\rmd{-\sigma\alpha(m_+)}{\alpha_1}}{\alpha_1}\right).
        \end{split}\end{equation}
        Furthermore, one can always find integers $a_\pm\in\mathbb{Z}$ such that $a_-\alpha_1-a_+\alpha_2=1$ (by virtue of Bezout's lemma). Then we can rewrite
        \begin{equation}
            m_{\pm}=(a_\pm+t\alpha_{1,2})\mathfrak{m}.
        \end{equation}
        Putting all this together, we arrive at
        \begin{equation}
            s=\frac{1}{2}\left(1+\mathfrak{b}-\mathfrak{c}-\chi_{1}-2\ii \alpha(a)\right)=-\frac{\alpha(\mathfrak{m})}{2\alpha_1\alpha_2}+\frac{\rmd{a_-\alpha(\mathfrak{m})}{\alpha_2}}{\alpha_2}-\ii\alpha( a),
        \end{equation}
        \begin{equation}
            t=\frac{\sigma}{2}\left(1-\mathfrak{b}-\mathfrak{c}-\chi_{1}-2\ii\alpha( a)\right)=1+\sigma\frac{\alpha(\mathfrak{m})}{2\alpha_1\alpha_2}-\frac{\rmd{-\sigma a_+\alpha(\mathfrak{m})}{\alpha_1}}{\alpha_1}-\ii\sigma\alpha( a).
        \end{equation}
        We are now ready to substitute into the one-loop determinant \eqref{eq.1loop.spindle}.
        
        \paragraph{Topological Twist.}
            This is the case of $\sigma=1$ and we find
            \begin{equation}\label{eq--Z.spindle.top}
                Z^{\text{top}}_{\mathbb{CP}^1_{\boldsymbol{\alpha}}}=\prod_{\alpha}\prod_{k=0}^\infty\frac{k-\frac{\alpha(\mathfrak{m})}{2\alpha_1\alpha_2}+\frac{\rmd{a_-\alpha(\mathfrak{m})}{\alpha_2}}{\alpha_2}-\ii\alpha( a)}{k+1+\frac{\alpha(\mathfrak{m})}{2\alpha_1\alpha_2}-\frac{\rmd{-a_+\alpha(\mathfrak{m})}{\alpha_1}}{\alpha_1}-\ii\alpha( a)}.
            \end{equation}
            We can rewrite this expression as follows
            \begin{equation}\begin{split}
                \alpha(\mathfrak{m})\geq 0:\; Z^{\text{top}}_{\mathbb{CP}^1_{\boldsymbol{\alpha}}}=&\prod_\alpha\prod_{k=0}^{\frac{\alpha(\mathfrak{m})}{\alpha_1\alpha_2}-\frac{\rmd{a_-\alpha(\mathfrak{m})}{\alpha_2}}{\alpha_2}-\frac{\rmd{-a_+\alpha(\mathfrak{m})}{\alpha_1}}{\alpha_1}}\left(k-\frac{\alpha(\mathfrak{m})}{2\alpha_1\alpha_2}+\frac{\rmd{a_-\alpha(\mathfrak{m})}{\alpha_2}}{\alpha_2}-\ii\alpha( a)\right)\\
                =&\prod_\alpha\prod_{j=-\frac{\alpha(\mathfrak{m})}{2\alpha_1\alpha_2}+\frac{\rmd{a_-\alpha(\mathfrak{m})}{\alpha_2}}{\alpha_2}}^{\frac{\alpha(\mathfrak{m})}{2\alpha_1\alpha_2}-\frac{\rmd{-a_+\alpha(\mathfrak{m})}{\alpha_1}}{\alpha_1}}\left(j-\ii\alpha( a)\right),
            \end{split}\end{equation}
            \begin{equation}\begin{split}
                \alpha(\mathfrak{m})< 0:\;Z^{\text{top}}_{\mathbb{CP}^1_{\boldsymbol{\alpha}}}=&\prod_\alpha\prod_{k=0}^{-\frac{\alpha(\mathfrak{m})}{\alpha_1\alpha_2}-2+\frac{\rmd{-a_+\alpha(\mathfrak{m})}{\alpha_1}}{\alpha_1}+\frac{\rmd{a_-\alpha(\mathfrak{m})}{\alpha_2}}{\alpha_2}}\left(k+1+\frac{\alpha(\mathfrak{m})}{2\alpha_1\alpha_2}-\frac{\rmd{-a_+\alpha(\mathfrak{m})}{\alpha_1}}{\alpha_1}-\ii\alpha( a)\right)^{-1}\\
                =&\prod_\alpha\prod_{l=\frac{\alpha(\mathfrak{m})}{2\alpha_1\alpha_2}-\frac{\rmd{-a_+\alpha(\mathfrak{m})}{\alpha_1}}{\alpha_1}+1}^{-\frac{\alpha(\mathfrak{m})}{2\alpha_1\alpha_2}-1+\frac{\rmd{a_-\alpha(\mathfrak{m})}{\alpha_2}}{\alpha_2}}\left(l-\ii\alpha( a)\right)^{-1},
            \end{split}\end{equation}
            where $j=k-\frac{\alpha(\mathfrak{m})}{2\alpha_1\alpha_2}+\frac{\rmd{a_-\alpha(\mathfrak{m})}{\alpha_2}}{\alpha_2}$ and $l=k+1+\frac{\alpha(\mathfrak{m})}{2\alpha_1\alpha_2}-\frac{\rmd{-a_+\alpha(\mathfrak{m})}{\alpha_1}}{\alpha_1}$. A further rewriting yields
            \begin{equation}\begin{split}\label{eq.1looptop.app}
                Z^{\text{top}}_{\mathbb{CP}^1_{\boldsymbol{\alpha}}}=&\prod_{\alpha>0}\prod_{j=-\frac{\alpha(\mathfrak{m})}{2\alpha_1\alpha_2}+\frac{\rmd{a_-\alpha(\mathfrak{m})}{\alpha_2}}{\alpha_2}}^{\frac{\alpha(\mathfrak{m})}{2\alpha_1\alpha_2}-\frac{\rmd{-a_+\alpha(\mathfrak{m})}{\alpha_1}}{\alpha_1}}\left(j-\ii\alpha( a)\right)\prod_{l=-\frac{\alpha(\mathfrak{m})}{2\alpha_1\alpha_2}-\frac{\rmd{+a_+\alpha(\mathfrak{m})}{\alpha_1}}{\alpha_1}+1}^{\frac{\alpha(\mathfrak{m})}{2\alpha_1\alpha_2}-1+\frac{\rmd{-a_-\alpha(\mathfrak{m})}{\alpha_2}}{\alpha_2}}\left(l+\ii\alpha( a)\right)^{-1}\\
                =&\prod_{\alpha>0}\prod_{j=-\frac{\alpha(\mathfrak{m})}{2\alpha_1\alpha_2}+\frac{\rmd{a_-\alpha(\mathfrak{m})}{\alpha_2}}{\alpha_2}}^{\frac{\alpha(\mathfrak{m})}{2\alpha_1\alpha_2}-\frac{\rmd{-a_+\alpha(\mathfrak{m})}{\alpha_1}}{\alpha_1}}\left(j-\ii\alpha( a)\right)\prod_{l=-\frac{\alpha(\mathfrak{m})}{2\alpha_1\alpha_2}-\frac{\rmd{-a_-\alpha(\mathfrak{m})}{\alpha_2}}{\alpha_2}+1}^{\frac{\alpha(\mathfrak{m})}{2\alpha_1\alpha_2}-1+\frac{\rmd{+a_+\alpha(\mathfrak{m})}{\alpha_1}}{\alpha_1}}\left(l-\ii\alpha( a)\right)^{-1}\\
                =&\prod_{\alpha>0}\left(\frac{\alpha(\mathfrak{m})}{2\alpha_1\alpha_2}+\ii\alpha( a)\right)\left(-\frac{\alpha(\mathfrak{m})}{2\alpha_1\alpha_2}+\ii\alpha( a)\right).
            \end{split}\end{equation}
            Here we used the relation \eqref{eq.relation.mmod} and, between the first and the second line, we neglected an overall minus sign in the second factor while also sending $l\rightarrow -l$. The first factor in the bottom line appears only if $\rmd{a_-\alpha(\mathfrak{m})}{\alpha_2}=\rmd{a_-\alpha(\mathfrak{m})}{\alpha_2}=0$ and the second only if $\rmd{-a_+\alpha(\mathfrak{m})}{\alpha_1}=\rmd{a_+\alpha(\mathfrak{m})}{\alpha_1}=0$. Employing \eqref{eq.condition.apm} these two conditions are equivalent, respectively, to $\alpha(\mathfrak{m})\in\alpha_1\mathbb{Z}$ and $\alpha(\mathfrak{m})\in\alpha_2\mathbb{Z}$. Finally, this agrees with our result \eqref{eq.1loop.spindle.top} on the branched cover upon setting $\epsilon_1=1$ and a constant shift in the Coulomb branch parameter in \eqref{eq.1loop.spindle.top},
            \begin{equation}\label{eq.shiftcoulomb}
                a\rightarrow a-\ii\frac{\alpha(\mathfrak{m})}{\alpha_1\alpha_2}.
            \end{equation}
            This constant shift is such that also the classical contribution to the partition function matches \cite{Lundin:2021zeb}.
        
        \paragraph{Exotic.}
            This is the case of $\sigma=-1$ and we find
            \begin{equation}\begin{split}
                Z^{\text{ex}}_{\mathbb{CP}^1_{\boldsymbol{\alpha}}}=&\prod_{\alpha}\prod_{k=0}^\infty\frac{k-\frac{\alpha(\mathfrak{m})}{2\alpha_1\alpha_2}+\frac{\rmd{a_-\alpha(\mathfrak{m})}{\alpha_2}}{\alpha_2}-\ii\alpha( a)}{k+1-\frac{\alpha(\mathfrak{m})}{2\alpha_1\alpha_2}-\frac{\rmd{a_+\alpha(\mathfrak{m})}{\alpha_1}}{\alpha_1}+\ii\alpha( a)}\\
                =&\prod_{\alpha>0}\prod_{k=0}^\infty\frac{k-\frac{\alpha(\mathfrak{m})}{2\alpha_1\alpha_2}+\frac{\rmd{a_-\alpha(\mathfrak{m})}{\alpha_2}}{\alpha_2}-\ii\alpha( a)}{k+1-\frac{\alpha(\mathfrak{m})}{2\alpha_1\alpha_2}-\frac{\rmd{a_+\alpha(\mathfrak{m})}{\alpha_1}}{\alpha_1}+\ii\alpha( a)}\frac{k+\frac{\alpha(\mathfrak{m})}{2\alpha_1\alpha_2}+\frac{\rmd{-a_-\alpha(\mathfrak{m})}{\alpha_2}}{\alpha_2}+\ii\alpha( a)}{k+1+\frac{\alpha(\mathfrak{m})}{2\alpha_1\alpha_2}-\frac{\rmd{-a_+\alpha(\mathfrak{m})}{\alpha_1}}{\alpha_1}-\ii\alpha( a)}\\
                =&\prod_{\alpha>0}\prod_{k=0}^{\frac{\alpha(\mathfrak{m})}{\alpha_1\alpha_2}-\frac{\rmd{a_-\alpha(\mathfrak{m})}{\alpha_2}}{\alpha_2}-\frac{\rmd{-a_+\alpha(\mathfrak{m})}{\alpha_1}}{\alpha_1}}\left(k-\frac{\alpha(\mathfrak{m})}{2\alpha_1\alpha_2}+\frac{\rmd{a_-\alpha(\mathfrak{m})}{\alpha_2}}{\alpha_2}-\ii\alpha( a)\right)\\
                \quad\quad&\prod_{k=0}^{\frac{\alpha(\mathfrak{m})}{\alpha_1\alpha_2}-2+\frac{\rmd{+a_+\alpha(\mathfrak{m})}{\alpha_1}}{\alpha_1}+\frac{\rmd{-a_-\alpha(\mathfrak{m})}{\alpha_2}}{\alpha_2}}\left(k+1-\frac{\alpha(\mathfrak{m})}{2\alpha_1\alpha_2}-\frac{\rmd{+a_+\alpha(\mathfrak{m})}{\alpha_1}}{\alpha_1}+\ii\alpha( a)\right)^{-1}\\
                =&\prod_{\alpha>0}\prod_{j=-\frac{\alpha(\mathfrak{m})}{2\alpha_1\alpha_2}+\frac{\rmd{a_-\alpha(\mathfrak{m})}{\alpha_2}}{\alpha_2}}^{\frac{\alpha(\mathfrak{m})}{2\alpha_1\alpha_2}-\frac{\rmd{-a_+\alpha(\mathfrak{m})}{\alpha_1}}{\alpha_1}}\left(j-\ii\alpha( a)\right)\prod_{l=-\frac{\alpha(\mathfrak{m})}{2\alpha_1\alpha_2}+1-\frac{\rmd{+a_+\alpha(\mathfrak{m})}{\alpha_1}}{\alpha_1}}^{\frac{\alpha(\mathfrak{m})}{2\alpha_1\alpha_2}-1+\frac{\rmd{-a_-\alpha(\mathfrak{m})}{\alpha_2}}{\alpha_2}}\left(l+\ii\alpha( a)\right)^{-1}
            \end{split}\end{equation}
            This is the expression we encountered in the previous case, which reduces to:
            \begin{equation}\label{eq.1loopex.app}
                Z^{\text{ex}}_{\mathbb{CP}^1_{\boldsymbol{\alpha}}}=\prod_{\alpha>0}\left(\frac{\alpha(\mathfrak{m})}{2\alpha_1\alpha_2}+\ii\alpha( a)\right)\left(-\frac{\alpha(\mathfrak{m})}{2\alpha_1\alpha_2}+\ii\alpha( a)\right),
            \end{equation}
            where the first term contributes only if $\alpha(\mathfrak{m})\in\alpha_1\mathbb{Z}$ and the second only if $\alpha(\mathfrak{m})\in\alpha_2\mathbb{Z}$. This agrees with \eqref{eq.1loop.spindle.ex} for $\epsilon_1=1$ and replacing $a\leftrightarrow\eta$. Also in this case the classical contributions can be shown to match what we found reducing from the branched cover.

\addtocontents{toc}{\protect\setcounter{tocdepth}{2}}

\bibliographystyle{utphys}
\bibliography{main}

\end{document}